\documentclass[useAMS,usenatbib]{mn2e}
\usepackage{graphicx}
\usepackage{txfonts}
\begin{document}

\title[Pulse broadening analysis for several pulsars]
{Pulse broadening analysis for several new pulsars and anomalous scattering}
\author[Wojciech Lewandowski et al.]
{Wojciech Lewandowski,\thanks{E-mail: boe@astro.ia.uz.zgora.pl}
Marta Dembska,
Jaroslaw Kijak,
Magdalena Kowali{\'n}ska\\
Kepler Institute of Astronomy, University of Zielona G\'ora , Lubuska~2, 65-265~Zielona~G\'ora, Poland}
\date{Accepted . Received ; in original form }
\maketitle

\begin{abstract}
We show the results of our analysis of the pulse broadening phenomenon in 25 pulsars at several frequencies using the data gathered with GMRT and Effelsberg radiotelescopes. Twenty two of these pulsars were not studied  in that regard before and our work has increased the total number of pulsars with multi-frequency scattering measurements to almost 50, basically doubling the amount available so far. The majority of  the pulsars we observed have high to very-high dispersion measures ($DM>200$) and our results confirm the suggestion of L{\"o}hmer~et~al.~(2001, 2004) that the scatter time spectral 
indices for high-$DM$ pulsars deviate from the value predicted by a single thin screen model with Kolmogorov's 
distribution of the density fluctuations. In this paper we discuss the possible explanations for such deviations.
\end{abstract}

\label{firstpage}

\begin{keywords}
{stars: pulsars -- general, pulsars -- scattering}
\end{keywords}

\section{Introduction}
 \label{intro}

The existence of the ionized fraction of the Interstellar Medium (ISM) affects the incoming pulsar radiation in several ways. The first order effect - the interstellar dispersion - is due to the amount of the free electrons along the line-of-sight, usually quantized by the Dispersion Measure: $DM =\int_0^D n_e dl$, where $D$ is the pulsar distance and $n_e$ is the electron density. This phenomenon causes the pulses to arrive at the observer with the greater delay, the lower the observing frequency is. This can also cause the so called ``dispersion smearing'' when using filterbank-type spectrometer receivers as there will be a non-zero delay between the pulse arrival at the upper and the lower end of the observing bandwidth (even for an individual spectral channel of the receiver used). The study of this phenomenon for the entire pulsar population, along with other types of observations (in both in radio and optical regimes), helped us to understand the distribution of free electrons in the Milky Way galaxy \citep{taylor93}. Apparently, the ionized medium in our galaxy consists of three main parts: giant H~II regions within the spiral arms, the thin and dense inner disk, and a diffuse outer disk.

Due to the turbulent nature of the interstellar medium and variations of the electron density along the line-of-sight, several other phenomena may also arise when observing pulsars. One of them is angular and temporal broadening, which are caused by the scattering of the radiation, so it reaches the observer after traveling along different geometrical paths. The other is the Interstellar Scintillation (ISS), which can be explained in terms of diffraction and refraction effects, and the interference of the pulsar radiation  wavefront - disturbed by the existence of non-uniform medium - with itself. The statistical analysis of the ISS helped us to understand the nature of the fluctuations present in the ISM \citep[see ][ for a review]{rickett90}.

In this paper we will focus on the scattering effects. Even if one assumes that the signal (a pulse) leaves the pulsar at one instant, then due to the scattering by the ISM it will reach the observer along different geometrical paths, with different lengths, the pulse will arrive at the observer over a finite interval, i.e. the pulse will be smeared and will attain a ``scattering tail''. The shape of the tail may be explained if the brightness
distribution of the scattered radiation has a two-dimensional gaussian form and the
scattering happens only within a thin screen. Doing so one can estimate that the pulse broadening function ($PBF$) will take a form of an exponential decay: $PBF(t) \sim \exp(t/\tau_d)$ where $\tau_d$ is called the scatter time \citep[or pulse broadening time, ][]{scheuer68}. This value will be strongly dependant of the observing frequency: $\tau_d \propto \nu^{-\alpha}$, where the spectral index $\alpha$ is the same as for the value of  the decorrelation bandwidth $\nu_d \propto \nu^{\ \alpha}$, which can be measured by the means of interstellar 
scintillation observations. Both these values are related as in:

\begin{equation}
\label{def_c}
2\pi\, \tau_d\, \Delta\nu_d = C_1,
\end{equation}

\noindent
where $C_1$ is assumed to be of the order of unity, although it may vary for different ISM geometries and/or models of the ISM fluctuations \citep{lambert99}. To estimate the strength of scattering effects one has to assume a reasonable spectrum of electron density fluctuations. The most common way \citep[following ][]{rickett77} is to assume a homogeneous isotropic turbulence that within a range of fluctuation scales between an {\it inner scale} $k_i^{-1}$ and the {\it outer scale} $k_o^{-1}$ has a power spectrum can be expressed as:

\begin{equation}
\label{elec_dens_full}
P_{n_e} (q) = \frac{C_{n_e}^2}{(q + k_o^2)\ ^{\beta/2}} \exp\left(-\frac{q^2}{4\ k_i^2}\right), 
\end{equation}

\noindent
where $q$ is an amplitude of a three-dimensional wavenumber and $C_{n_e}^2$ is the fluctuation strength for a given line-of-sight.

If the fluctuations causing the observed refractive/diffractive effects are indeed within that range (i.e. $k_o \ll q \ll k_i$) this equation simplifies into a simple power-law:

\begin{equation}
\label{elec_dens_simple}
P_{n_e} (q) = C_{n_e}^2 q^{-\beta},
\end{equation}

\noindent 
and if $\beta$ is lower than 4 one can derive that the scatter time spectral index $\alpha = 2\beta/(\beta -2)$. For a purely Kolmogorov spectrum of the density irregularities (one has to note that it was devised for a neutral gas) the spectral index $\beta = 11/3$ which yields the expected value of $\alpha = 4.4$.

It has to be mentioned however, that this simplification of Equation~\ref{elec_dens_full} to 
Eq.~\ref{elec_dens_simple} may not be valid for the real case scenarios, i.e. the inner/outer scale condition
 ($k_o \ll q \ll k_i$) will not be fulfilled. The outer scale is less of a problem, since its influence 
would make the spectra steeper than Kolmogorov, but as \citet{rickett09} pointed out, the hypothesis of 
a steeper spectrum was proposed, tested and found to be incorrect. Thus the spectral exponent must 
be of 4.4 or flatter. From that one can infer that the outer scale is indeed not a factor in scattering, and one
can safely assume that $k_o = 0$, simplifying Eq.~\ref{elec_dens_full}. 

On the other hand the inner scale influence is important, and the evidence for that
can be found in the pulse shape measurements at meter wavelengths. In general that means, that it
does not make sense to analyze observations of high $DM$ pulsars at low frequencies in terms of 
Eq.~\ref{elec_dens_simple}. When the (frequency dependent) diffractive scale falls below the 
inner scale, the phase structure function becomes quadratic, and the resulting spectral slope becomes 
$\alpha=4.0$. For an individual source this can result in a Kolmogorov slope ($\alpha=4.4$) at high observing 
frequencies, while at low frequencies the influence of the inner scale effects causes it to drop to 4.0.
A very detailed modern discussion of the pulse shape analysis and the possibility of the inner 
scale estimation is given in \citet{rickett09}.

Also, one has to remember that the  theory presented above assumes an isotropic and homogeneous turbulence. As it
was shown by \citet{rickett09} the anisotropy will not change the frequency scaling of the scattering directly, 
but it may significantly affect the pulse shape, causing the scattering tail to decay with two different 
timescales. This may cause confusing results when trying to approach such profile with modeling, and affect
the results obtained by applying a simple single-decay models. The inhomogeneity on the other hand may cause the 
scattering parameters to vary with time, as it was shown for low-$DM$ pulsars 
\citep[see][an the references therein]{brisken10}. Also, for high $DM$ pulsars the scattering may vary, even if the 
turbulence is homogeneous. All these effects may affect the observed spectral slope $\alpha$ if the observations 
at different frequencies are performed at different epochs, and may cause the slope of the frequency dependence 
to fall below 4.0, which is impossible in the homogeneous turbulence theory \citep{romani86}.

Since the discovery of the ISM effects in the observed pulsar radiation various authors tried to measure and analyze the properties of that interaction, and the attempts to estimate the scatter time spectrum (or the decorrelation bandwidth spectrum) are one of the most important, since it would allow us to estimate the properties of the density fluctuations in the ionized interstellar gas as the observed spectral index $\alpha$ (with the exceptions given above) gives us a direct access to the spectrum of the density fluctuations. The most complete analysis so far was made by \citet{L04} and their earlier paper, \citet{L01} - L04 and L01 hereafter. In L04 the authors gathered the measurements of the spectral index $\alpha$ - from both the scatter time and decorrelation bandwidth - from the literature \citep[][ and L01]{cordes85,johnston98,kuzmin02} and after adding their own observations they were able to obtain the spectral index for a total number of 27 pulsars. Probably the most important outcome of L01 and L04 was the observation that for pulsars with low dispersion measure ($DM <300$~pc~cm$^{-3}$) the spectral index $\alpha$ stays in a good agreement with the Kolmogorov's theory predictions (although the authors noted a few peculiarities for a few of the nearby pulsars) but for pulsars with large $DM$, i.e. the most distant ones in their sample, the value of $\alpha$ deviates significantly from the predicted value of $4.4$, making the spectra much flatter than expected (down to $\alpha = 3$, with an average of $\alpha = 3.44$). The authors attributed this discrepancy mainly to the fact that for the most distant sources the probability of multiple scatterings along the line-of-sight significantly increases and such occurrence would make any predictions based on a single thin screen model inappropriate.

In this paper we present our results of the scatter time measurements and the estimation of the spectral index $\alpha$  for 25 pulsars, 22 of which were never previously analysed in this regard. In combination with the results of L01 and L04 we will try to interpret the observed trends and discuss the possible reasons behind the deviations from the values predicted by the commonly used simplified\footnote{single thin screen with Kolmogorov's spectrum of density fluctuations} model; deviations which we also noted for some of the pulsars from our sample. One has to also signify that the total number of pulsars which we study in this paper is almost twice as big as the sample that was available for L04.

\section{Observations and Data Analysis}

The results of the measurements which are presented in this paper were collected over the course of another observing project, namely the search for gigahertz peaked spectra (GPS) pulsars \citep{kijak11}. That project involved the measurement of pulsar flux densities at several frequencies, using the Giant Meterwave Radio Telescope (GMRT, Pune, India) for the low-frequency observations at 325~MHz, 610~MHz and 1060~MHz, as well as the 100-meter Effelsberg Radiotelescope for the 2.6~GHz, 4.85~GHz and 8.35~GHz observations. The observations were conducted between 2005 and 2010.

\subsection{Observations}
\label{obs}
The GMRT observations were conducted with 16~MHz bandwidth with the use of the telescope's 
phased array mode, with a sampling rate of 0.512~milliseconds, and 256 spectral channels across the 
band \citep[see ][ for the receiving system details]{guptaetal00}. Since the original goal 
of the project was to measure pulsar flux densities, the observations were 
intensity-calibrated using measurements of known continuous sources (like 1822-096). The typical
integration times were between 20 and 30 minutes, up to 1 hour in some selected cases (i.e. 
very weak sources). More details about the specifics of our GMRT observations can be found in
\citet{kijak07}.

For the high frequency observations conducted with the 100-m radio telescope of the Max-Planck Institute for Radioastronomy  at Effelsberg we used the bandwidths of 100~MHz (at 2.6~GHz), 500~MHz (at 4.85) and 1.1 GHz (at 8.35~GHz). We used secondary focus receivers equipped with cooled HEMT amplifiers, providing LHC and RHC signals that were digitised and independently sampled at the rate of 1024 bins per pulse period and synchronously folded using the topocentric pulsar rotational period \citep{jessner96}. Again, the typical integration time was ca. 30 minutes \citep[see also ][]{kijak11}. The observations were intensity-calibrated by the use of a noise diode and known continuum radio-sources (NGC 7027, 3C 273, 3C 286). 

We found the sensitivity of our observations to be very good \citep[see the Appendix of this paper and also Table~1 in ][]{kijak11}, ranging from ca. 1~mJy at 325~MHz, to an average of 0.5~mJy at 610~MHz and even slightly better at 1060/1170~MHz. For the higher frequency observations with the use of the 100-meter Effelsberg Radiotelescope we reached the sensitivity of c.a. 0.1~mJy at 2.6~GHz (and much better for some selected sources), to $\approx 0.02$~mJy at 4.85~GHz, and close to 0.01~mJy at 8.35~GHz. In most of the cases such sensitivity is good enough not only to measure the pulsars flux densities but also allows us to reliably study the profile shapes of the detected pulsars.

The original goal of the project was to find pulsars with a high-frequency turn over in the spectrum. Following the suggestion of \citet{kijak07} that this phenomenon may be correlated with the pulsar dispersion measure, we have chosen a sample of sources with mid- to high-dispersion measures (with a few exceptions), to estimate their flux densities and construct the spectra of these pulsars over a wide range of frequencies. The results of the spectral aspect of these observations were published by \citet{kijak11}. Over the course of our analysis we realised however that the data we gathered may be used for other purposes as well. Specifically, one of the difficulties we encountered while trying to conduct the flux density measurements was the fact that the pulsar profiles we obtained were (at low frequencies) heavily affected by the interstellar scattering. The long scattering tails we detected on some of the pulsar profiles made the flux density measurements difficult. The flux measurement procedure requires finding of the profile background, and the long scattering tails - extending up to 300~degrees in pulsar phase from the main pulse - can make the estimation of the background level almost impossible.
\begin{figure}
\resizebox{\hsize}{!}{\includegraphics{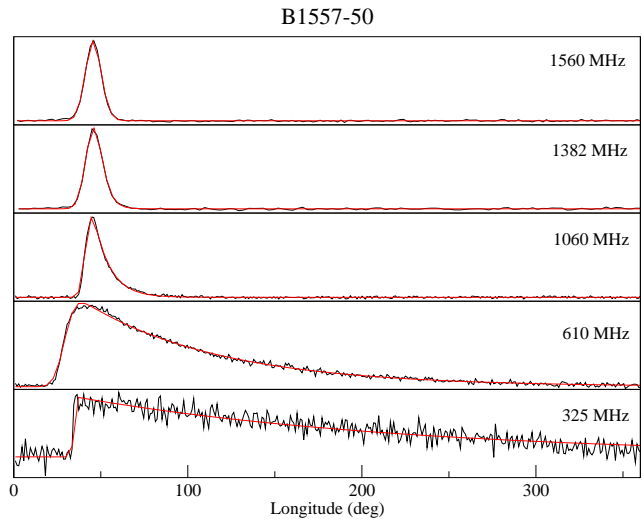}}
\caption{Profiles of B1557$-$50 at five frequencies, obtained with GMRT and Effelsberg. The apparent sharpness of the 325~MHz profile rise is probably caused by random noise variations and very long scatter time, which makes the scattering tail to contribute to (and distort) the next pulse (see article text).\label{1557_prof}}
\end{figure}

\begin{figure}
\resizebox{\hsize}{!}{\includegraphics{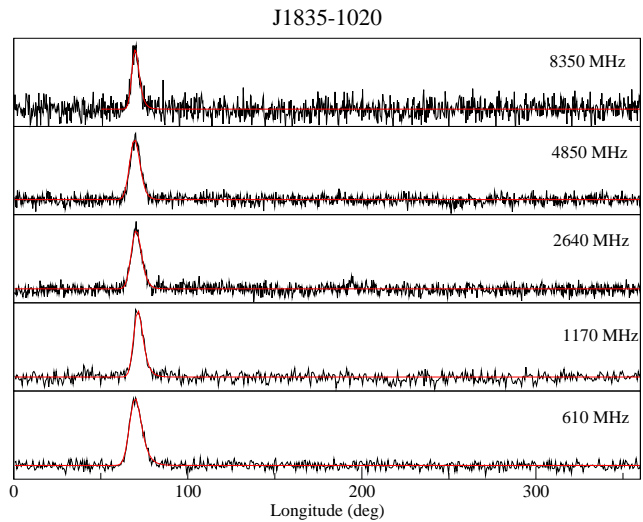}}
\caption{Profiles of J1835$-$1020 at five frequencies, obtained with GMRT and Effelsberg.\label{1835_prof}}
\end{figure}

It was then that we realised that the data we gathered can be also used to measure the scattering effect itself. It occurred to us as well that the sample of pulsars we have chosen was the perfect extension of the work done by L01 and L04 (see also Section~\ref{intro}). Since our sample included mostly the pulsars that had only very few flux density measurements, naturally the majority of them did not have their scattering properties analysed and certainly not at the broad range of observing frequencies that would allow the analysis of the scattering vs. frequency dependence. Thus we decided to analyse the data to measure the scattering effect for the pulsars in our sample.

Using the obtained pulse profiles we were also able to measure other profile parameters for the pulsars from our sample at several observing frequencies (namely, the $W_{50}$ and $W_{10}$ pulse widths). These measurements, along with the classification of the pulsar profiles \citep[see ][]{rankin83}, the flux density measurements for several pulsars (since not all of them were published before), and the profiles themselves can be found in the Appendix to this paper. Our scattering results were also partially published already - see \citet{kowal13}.

To supplement our observations we also used some of the archived profiles from the databases available in the internet, namely the EPN Database\footnote{\tt http://www.jb.man.ac.uk/pulsar/Resources/epn/} and the ATNF Database\footnote{The ANDSATNF database, available at \\ \tt http://www.atnf.csiro.au/research/pulsar/}, which were mainly used to obtain 1.4~GHz and 1.6~GHz profiles.

Two sample sets of the profiles we obtained can be seen in Figures~\ref{1557_prof} and \ref{1835_prof}, showing the aligned profile shapes at 5 frequencies each. PSR~B1557$-$50, shown in Fig.~\ref{1557_prof}, is clearly showing a strong scattering effect, especially at the lower frequencies. Despite the fact that the highest frequency shown is only 1560~MHz, and the lowest is 325~MHz, the shape of the profile changes from a gaussian-like with no detectable scattering tail to a smeared one with the tail extending over the whole 360 degrees. In fact the scatter time is in this case larger than the pulse period (see Section~\ref{errors}). This, along with the noise in the profile, causes the apparent sharpness of the profile rise, as the slope of the scattering tail ``from the previous pulse'' probably distorts the wings of the supposedly gaussian-like rise of the pulse, making it to appear sharper. This is probably also the reason why the baseline just prior to the pulse looks almost flat, while with a scatter time so big it should be sloped. The opposite is the case of PSR~J1835$-$1020, shown in Figure~\ref{1835_prof}, where almost no shape change can be seen by eye and yet the range of the observing frequencies at which those profiles were obtained is much greater (610~MHz to 8.35~GHz). Of course, the scattering effect depends strongly on the observing frequency, and the fact that the latter source was not observed at 325~MHz may ``smear'' the picture. The other source of this striking difference may be the fact that PSR~B1557$-$50 has a $DM$ of 260~pc~cm$^{-3}$, and for PSR~J1835$-$1020 the dispersion is lower by a factor of 2.5 ($DM = 113$~pc~cm$^{-3}$). Also, the goal of our project was to estimate the properties of the scattering phenomenon not only qualitatively but also quantitatively.

It has to be noted, that due to the nature of the original observing project, the multi-frequency observations were
not conducted simultaneously. While usually for a single source the individual low-frequency observations (performed at GMRT) were separated by a few days (up to two weeks), and the same stands for the high-frequency observations (at the Effelsberg telescope), the separation between the low and high-frequency observations epochs may be of the order of a few months, or even years. Additionally, some of the profiles from the EPN database we used to supplement our sample were conducted in 1990's, several years before our project started. This has to be taken into account when trying to interpret the results.

\subsection{Data Analysis and Results}

To estimate the strength of the scattering phenomenon in the obtained profiles, we used the method developed by L01, L04. In general, the observed pulse profile is the convolution of the intrinsic profile shape $P^I(t)$ with:

\begin{itemize}
\item the scattering smearing function $s(t)$, 
\item the dispersion smearing function $d(t)$; this function describes the smearing by the 
interstellar dispersion of the signal, which is present here due to a finite bandwidth 
of the filterbank used
\item the impulse response of the backend used $i(t)$.
\end{itemize}
Thus, the observed profile shape $P^O$ may be obtained as \citep{rama97}:

\begin{equation}
\label{profile_shape}
P^O = P^I(t) *  s(t) *  d(t) * i(t),
\end{equation}

\noindent
where the asterisks ($*$) denote the convolution.

The response times of the modern receiver systems are fast enough that we can assume the effect of $i(t)$ to be negligible. The dispersion smearing function $d(t)$ is in general a rectangular function in the case of an incoherent dedispersion (which was the case for the receiver systems we used). As it was mentioned before, for the low-frequency observations we were using the GMRT 256-channel filterbank system with 16~MHz bandwidth, which yields the channel width of 62.5~kHz. For a pulsar with $DM = 100$~pc~cm$^{-3}$ it means the dispersion smearing is of the order of 1.5~millisecond, while it is naturally 5 times larger for a $DM = 500$ pulsar (7.5~ms). With the 0.512~ms sampling rate we used, those values  mean, that for most of the pulsars from our sample (mid- to high-dispersion measure) this effect will introduce a smearing of between a few up to ca. 20 samples in the obtained profiles. For a typical 0.5 second pulsar this corresponds to 1-2\% of a pulse window and thus the dispersion smearing will significantly affect the observed shape of the profile (since for most pulsars its intrinsic width is usually around a few percent). This was not a problem for our original project goal (the flux density measurements), since only the shape of the profile is changed but the amount of energy integrated over a broad-enough pulse window stays constant. For the purposes of the scattering measurements however one has to take it into consideration.

While our low-frequency observations may have suffered a significant dispersion smearing due to a fact that this effect is stronger at lower frequencies, one has to remember about it at the higher frequencies as well. While the dispersion gets weaker, the high observing frequency filterbank receivers tend to use much wider bandwidths. Since our initial goal was not dependant on the profile shape, and we only wanted to make sure that we measure all the incoming energy, for the observations with the 100-meter Effelsberg Radiotelescope we used the maximum available bandwidths, even at the cost of not being able to dedisperse the data. This means that our high frequency observations will suffer the dispersion smearing due to wide bandwidths, i.e. 500~MHz at 4.85~GHz and 1.1~GHz bandwidth at 8.35~GHz. Again, for a $DM=100$ pulsar this translates to a dispersion smearing of $\approx 4$~millisecond, or 20~ms (ca. 4\% of the cycle) for a pulsar with $DM=500$ (and a period of 0.5 seconds).

The $s(t)$ component in Eq.~\ref{profile_shape} describes the broadening of a pulse due to the interstellar scattering. Different mathematical forms were used to model this effect but as it was discussed by L04 a simple exponential decay function seems to work best for highly dispersed pulsars (which was the case for most of the objects in our sample); a function that describes scattering on a single thin screen \citep[see ][]{williamson72, williamson73}. In our analysis we assumed that the scatter broadening $s(t)$ has a form of $PBF_1$ function from L04:

\begin{equation}
\label{pbf1}
PBF_1 = \exp(-t/\tau_d) U(t),
\end{equation}

\noindent
where $\tau_d$ is the pulse broadening time, and $U(t)$ is the unit step function ($U(t<0)=0$, and $U(t \geq 0) = 1$).
 
To find the values of $\tau_d$ for the observed profiles we performed a least squares fit of a model shape to the observational data, to find a best fit parameters by minimizing the $\chi^2$ function (L01):

\begin{equation}
\label{chi2}
\chi^2 = \frac{1}{(N-4)\sigma^2_{\mbox{\scriptsize off}}} \sum_{i-1}^n \ \left[ P_i^O - P_i^M \ (a,b,c,\tau_d)\right]^2,
\end{equation}

\noindent
where $\sigma_{\mbox{\scriptsize off}}$ is the off-pulse rms, $N$ is the number of bins on the observed profile $P^O$ is the observed profile, and the $P^M$ is the model profile which besides the scatter time $\tau_d$ has additional three free parameters: the amplitude scaling factor $a$, the profile background level $b$, and the phase $c$.

The fit itself was performed using the {\it Origin} software\footnote{http://www.originlab.com/}. Since we do not know the intrinsic pulse profile (this is in general impossible to get) we used a convolution of a simple gaussian profile with the exponential decay due to the scatter broadening. This may be viewed as a simplified approach, but on the other hand - since we already had to deal with the dispersion broadening (due to the very low observing frequencies or wide bandwidths used at high frequencies, see the discussion above) - we decided that it was well justified. We were hesitant to use high frequency profiles as templates, as they were dispersion-broadened as well, and additionally one can not exclude the possible effects of intrinsic profile changes with the observing frequency (like profile evolution and radius-to-frequency mapping). These effects would ruin any attempts to use the unscattered high frequency profiles as the templates for the low-frequency scatter-broadening fit. One has to add however that our simplified approach introduces an additional parameter to the model profile $P^M$, i.e. the (half)width of the gaussian profile.

In the cases of a clearly multi-component profiles (i.e. double ones, like J1809$-$1917 in the appendix) the fit was usually performed to the second (trailing) component only. We used this only when the profile components were clearly separated, which would ensure that the scattering tail of the leading component would add only a negligible contribution to the tail of the trailing component, that was used for the actual fit.

The error estimates for the model parameters (namely $\tau_d$) were obtained by the method of Chi-square mapping, from the 1-$\sigma$ contours on the $c$-$\tau_d$ plane.

Table~\ref{tau_table} presents our measurements of the scatter time $\tau_d$ for our observations, as well as the results we obtained for profiles that were found in various databases available in the web, that we used to supplement our observations (for better frequency coverage). The quality of our fits can be also observed in Figures~\ref{1557_prof} and~\ref{1835_prof}, where the red line represents the result of the best
least squares fit to the observed profiles.

\begin{table*}
\caption{Scattering timescale $\tau_{sc}$ versus observing frequency\label{tau_table}. The values quoted in italic are considered doubtful, due to either poor quality of the profile (or the fit), or the fact that the value does not seem to match the frequency evolution of the scatter time (see Sections~\ref{errors} and~\ref{section_evolution} for details).}

\begin{tabular}{lcccccccc}
\hline
& \multicolumn{7}{c}{$\tau_{sc}$} \\ \cline{2-9}
Pulsar\phantom{\huge X} & 325 MHz & 610 MHz & 1060 MHz & 1.4 GHz & 1.6 GHz & 2.6 GHz & 4.85 GHz & 8.35 GHz \\ \hline \hline
B1557$-$50      &411$\pm150$ & 43$\pm3$&5.0$\pm0.6$ &1.98$\pm0.35$ & 0.97$\pm0.24$& --& --&--\\ 
 B1641$-$45     & --& 73.4$\pm4.8$&8.8$\pm0.4$ &-- &-- &--& {\it 2.5$\pm$0.4} &--\\ 
 J1740+1000     & --& 8.0$\pm3.5$& 1.6$\pm0.3$$^\star$& --& --& --& --&--\\ 
 B1740$-$31     & 252$\pm23$& 24.8$\pm5.5$& 1.1$\pm0.3$& --&-- &-- &-- &--\\ 
 B1750$-$24     & --& --& --& 85.6$\pm3.9$& 43.8$\pm2.3$& 13.6$\pm$4.2 & --&--\\ 
J1751$-$3323    & --& --& --& 24.4$\pm4.1$& --& 0.8$\pm0.6$&-- &--\\ 
 B1758$-$23     & --&-- &-- &99$\pm19$ &51$\pm10$ &{\it 0.75$\pm$0.34} &0.23$\pm0.08$ &--\\ 
 B1800$-$21     &-- &{\it 39$\pm$11} &0.22$\pm0.02$$^\dagger$ &0.06$\pm0.03$ &-- &-- &-- &--\\ 
 J1809$-$1917   & --& --& 2.9$\pm1.1$$^\star$&2.1$\pm0.8$ &-- & {\it 0.03$\pm$0.02}& --&--\\ 
 B1815$-$14     & --& 457$\pm150$& 44$\pm12$& 15.3$\pm2.0$& 8.7$\pm0.9$&-- &-- &--\\ 
 B1820$-$14     & --& 56.1$\pm3.7$& 5.6$\pm1.9$& 2.3$\pm0.8$& --&-- &-- &--\\ 
 B1822$-$14     & --& 143$\pm31$& 15.1$\pm2.0$& 6.1$\pm1.2$& 3.7$\pm1.5$& --& --&--\\ 
 B1823$-$11     & 218.8$\pm44$& 33$\pm13$& --& --& --& --&-- &--\\ 
  B1823$-$13    & --& --& 3.2$\pm1.6$$^\dagger$,3.7$\pm1.5$& 3.0$\pm0.5$& 1.8$\pm0.3$&-- &-- &--\\ 
 B1828$-$11     & --& 4.7$\pm0.4$& --& 0.45$\pm0.4$& --& 0.04$\pm0.04$& {\it 0.03$\pm$0.03}&--\\ 
  B1830$-$08    & --& --& 4.1$\pm1.3$$^\dagger$&2.6$\pm0.8$ &-- &-- &0.12$\pm0.06$ &--\\ 
 B1832$-$06     & --& --& 91$\pm9$& 26$\pm5$& 12.8$\pm4.0$&{\it 0.3$\pm$0.1} &0.12$\pm0.05$ &--\\ 
  B1834$-$04    & --& 33$\pm5$& 0.15$\pm0.10$& --&-- &-- &-- &--\\ 
 B1838$-$04     & --& 23$\pm7$& 2.4$\pm0.9$& 0.9$\pm0.2$& --&-- &-- &--\\ 
  J1835$-$1020  & --& 2.55$\pm0.34$& 2.08$\pm0.41$$^\star$&-- & --& 1.7$\pm0.2$& 1.7$\pm0.4$&1.6$\pm0.4$\\
J1842$-$0905    & --& 9.0$\pm1.0$& --& 0.11$\pm0.05$& --& --& --&--\\
 J1857+0143     & --& --& 24$\pm5$$^\star$& --& --&{\it 0.05$\pm$0.01} & 0.03$\pm0.02$&--\\
J1901+0510      & --& 10$\pm9$& --& 0.42$\pm0.09$& --& --& --&--\\
 J1907+0918     & -- & 3.6$\pm0.3$&0.28$\pm0.03$$^\star$ &-- & --& \it 0.06$\pm$0.01&-- &--\\
J1910+0728      & --& 10.2$\pm1.5$& --& 1.2$\pm0.2$& --& --& --&--\\ \hline
\multicolumn{8}{l}{Remarks: $^\dagger$ - 925 MHz,$^\star$ - 1170~MHz} \\
\end{tabular}
\end{table*}

\subsection{Possible sources of errors in the $\tau_d$ estimates}
\label{errors}
One has to ask a question: how reliable are the $\tau_d$ measurements using the above described method? There are at least a few factors that can influence the results.

First of them is the fact that we modelled all of our profiles with a gaussian function, which in case of most of the pulsars is not true. We believe that this issue will be somewhat alleviated by the dispersion smearing present in our data (as it was discussed above). As most of the objects in our sample are mid- to high-dispersion pulsars, their profiles will be severely smeared, both at the lowest and the highest frequencies that they were observed at. 

We believe that in the case of low-frequency observations, where the pulsars scattering tails are very long, the effect of the gaussian-approximation of the profile on the measured value of $\tau_d$ will be negligible. The majority of the points that are used for the actual fit, i.e. the points that contribute the most to the $\chi^2$ value that we ought to minimize, are lying far from the actual (unscattered) profile. This means that any profile details that may contribute to the scattering tail shape will be naturally smeared by the scattering effect itself, and adding dispersion smearing on top of that makes the influence of the intrinsic pulse shape even more insignificant.

On the other hand, at the high frequencies one has to analyse and review the obtained values carefully. Since we used broadband receiving systems and the resulting dispersion smearing may be of the order of the intrinsic pulse width in the extreme cases. For example a $DM=500$~pc~cm$^{-3}$ pulsar with a period of 0.5 seconds and 5\% duty cycle of the intrinsic profile would still have 5~ms smearing, which equals to 20\% of the pulse width. It is clear that the profile would be severily smeared but it would still retain the most important characteristics, like the profile asymetry. Blindly using our method for such pulsars may result in obtaining erroneous values. The most severe error would be made in the case, when the intrinsic profile is asymetric with the trailing part of the profile descending with a lower slope that would ``simulate'' a scattering tail. Adding the dispersion smearing and using our fit to such profile would result in a fake value of $\tau_d$ since the actual scatter-broadening would be neglible at high observing frequency. For this reason we have disregarded a few of the values presented in Table~\ref{tau_table} at the highest frequencies (2.6~GHz, 4.85~GHz and 8.35~GHz) in our further analysis - see Section~\ref{section_evolution} for discussion of individual cases.

\begin{figure*}
\resizebox{\hsize}{!}{\includegraphics{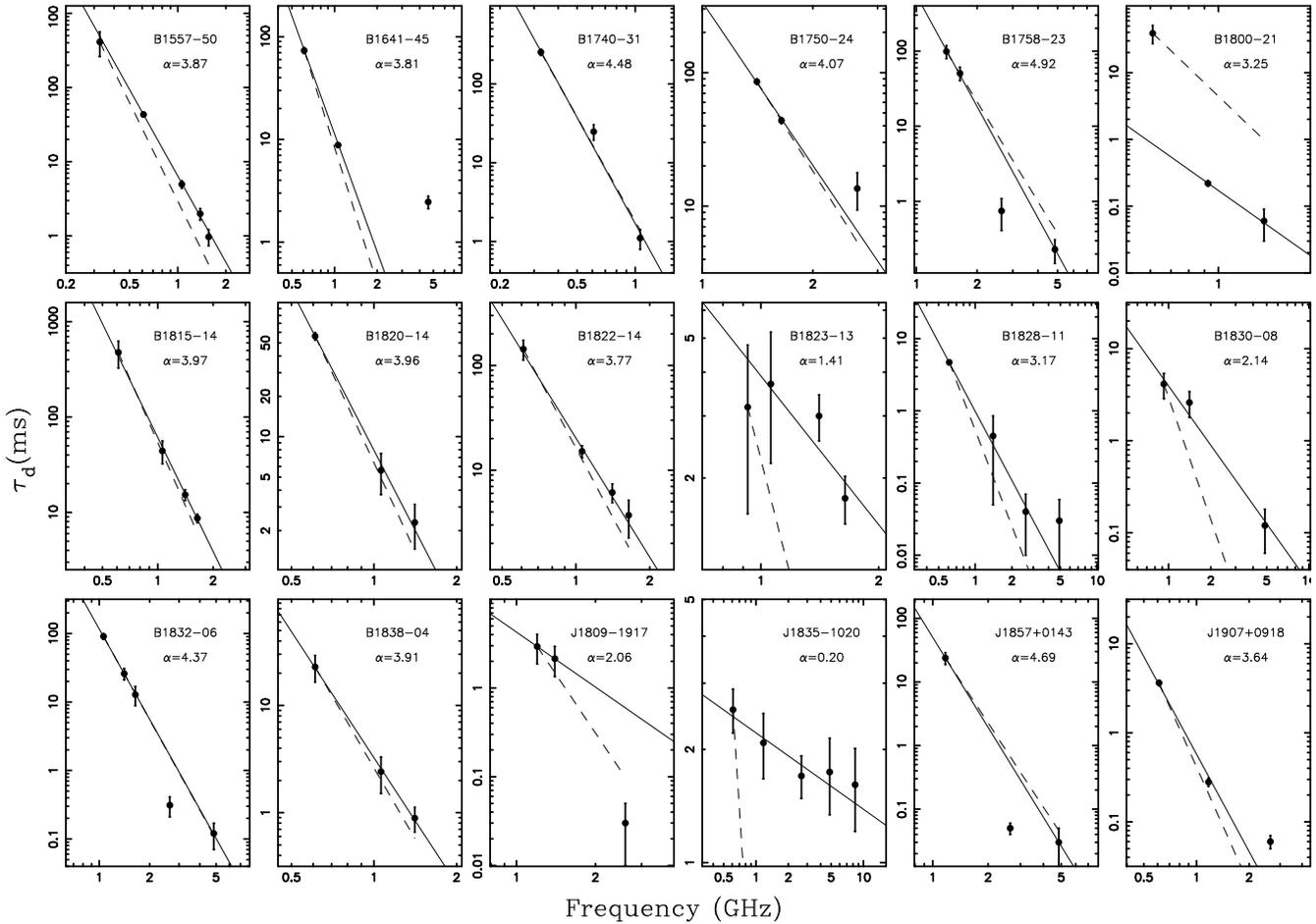}}
\caption{A plot of $\tau_{d}$ vs the observing frequency for pulsars from our sample. 
The solid line on each plot is the representation of the power-law fit to the scatter time measurements, while the dashed line shows the slope of the thin-screen Kolmogorov prediction with $\alpha = 4.4$. Note that some of the data points were omitted
in the actual fits - see Table~\ref{tau_table}, and the explanations in Sections~\ref{errors} and~\ref{section_evolution}. 
\label{plot_tau_vs_freq}}
\end{figure*}

The worst case scenario happens usually at the intermediate observing frequencies, where the intrinsic pulsar profiles, the dispersion smearing and the actual scattering have similar timescales. One can not negate the possibility that the values measured by fitting our simplified model may be affected by the non-gaussian shape of the intrinsic profile, and the measured value of the scatter time may be both higher than the actual value (in cases of pulse shapes similar to the one discussed in the previous paragraph) but it may also be lower. For the later case it is easy to imagine a profile with a low slope leading edge of the pulse and a sharp trailing edge. Adding a significant scattering to such profile may lead to evening the slopes on both sides of the profile maximum. Modeling such profile with a gaussian$+$exponential tail convolution would result in fitting just a broad gaussian to the profile with very little scattering (if any), thus the obtained value of $\tau_d$ would be much lower than the actual value.

This shows that in every case the fit has to be done very carefully, and one has to decide for each individual value obtained from the fit if it is reliable; a comparison of profiles from different observing frequencies (even if they are all affected by dispersion smearing) usually can help with that.

Of course it would be the best if we could eliminate at least some factors affecting the observed profiles, especially the dispersion smearing - this would allow us to use the pulse shapes from higher frequencies as intrinsic profile templates. This would be possible in case of a coherent dedispersion system, which eliminated the dispersion smearing completely. One can also at least try to minimize the effect of the dispersion smearing (at higher observing frequencies) by using narrower bandwidths. We intend to do so for the future observations of the scattering in pulsars but for the data we had available for this project it was simply not possible.

Another source of errors may be the scattering phenomenon itself. For highly scattered profiles, where the scatter time is comparable to (or even greater than) the pulsar period, it is still possible to see pulsed emission, but it is impossible to find the proper baseline of the profile. If, say, the scatter time is equal to the pulsar period, the scattering tail will still contribute $1/e$ of the ``real'' pulse amplitude after 360$^{\circ}$ of phase, i.e. to the next pulse, and of course also to that part of the profile which one would consider to be the profile baseline, effectively rising (and adding a bit of slope) to the observed baseline. When trying to fit the scattered pulse convolution profile without the correction of the baseline, one may still find a satisfactory solution, but the incorrect assumption that one sees the actual profile baseline (and the end of the scattering tail) just before the profile maximum, will of course limit the measured value of the scatter time to less than one pulsar period. 

It is generally impossible to resolve this problem for an individual highly scattered profile, since one needs to know how much was the  baseline raised to estimate the real scatter time, but to do that one needs the value of the scatter time. To at least partially overcome this problem we used the fact that we had profiles at multiple frequencies available. One of such cases was the pulsar B1557$-$50 (see Figure~\ref{1557_prof}). We used the profile from 610~MHz and found the scattering at this frequency to be of the order of 80~degrees (43~ms for a pulsar with 193~ms period). Using this value, and assuming a reasonable scatter time frequency evolution (see Section~\ref{section_evolution}, we tried the spectral slopes $\alpha$ between 4.0 and 4.4), we estimated the scatter time at the frequency of 325~MHz to be of the order of 1000 degrees. We then adjusted the baseline accordingly, and performed the scattered profile fits. The best fit results we got for  the actual scatter time of about 750 degrees (411~ms quoted in Table~\ref{tau_table}). This is of course very crude estimate (which is somewhat reflected in the estimated errors of the order of 30\%), but we find it much more plausible than the results of the direct fit to the profile with un-corrected baseline, which yields $\tau_sc \approx 100$~ms (i.e ca. 200 degrees, when the ``expected'' value is about 1000 degrees).

PSR~B1557-50 was the most extreme case we had to use this method, and we find it amazing that for a pulsar with the scatter time of the order of two periods we are still able to see pulsed emission. We had to use it also for 
PSR~B1815$-$14 and PSR~B1822$-$14. In both cases we used the data from 1060~MHz to estimate the baseline correction at 610~MHz, and doing so yielded the scatter times of the order of 550$^{\circ}$ and 185$^{\circ}$ respectively. For the remaining pulsars the lowest observed frequency scattering is much less than 120~degrees, and the baseline correction is negligibly small.

As one can see, a careful analysis of each individual profile was the only possibility for our analysis. In some cases, as it will be discussed in next section, the fact that we conducted multi-frequency observations was also very helpful with the interpretation and the reliability check of our results.

\section{The frequency evolution of pulse broadening}
\label{section_evolution}

In the previous section we have shown the results of our measurement of the scatter time $\tau_d$ from our observations, as well as from the archived profiles for the pulsars included in our sample, which we used to supplement our observations. Figure~\ref{plot_tau_vs_freq} shows these results in a graphical form, with each panel representing one pulsar. In the figure we included only the sources that had measurements of $\tau_d$ at three or more observing frequencies (18 of 25 pulsars that we analyzed). 

As it was mentioned in the introduction, we expect the measured values to be dependant on the observing frequency. Using Kolmogorov's distribution of the spatial density irregularities, assuming a single thin screen model (or a thick medium), and also that the {\it inner scale} of the density fluctuations is negligibly small, then the expected variation of the scatter time over any range of observing frequencies would be a simple power-law function, with a spectral index $\alpha = 4.4$. However, it has to be mentioned that for other models of the ISM distribution, or when the fluctuation {\it inner scale} is larger, one can get the expected $\tau_d$ spectral index of $4.0$, or lower \citep[see for example ][]{cordes01} - even with the  assumption of the Kolmogorov's irregularities distribution within ISM.

\begin{table*}
\caption{The scattering spectral index ($\alpha$)and the electron density spectral index ($\beta$) for the observed pulsars. Values of $\alpha$ and $\beta$ quoted in italic are considered doubtful (see article text for explanation). Values of $\beta$ indicated by an asterisk are ``unphysical'', as the $\beta=2\alpha/(\alpha-2)$ relation is valid only for $\alpha \geq 4$.   \label{table_alpha}}
\begin{tabular}{lrrrrllc}\hline
Pulsar & $l_{II}~~$ &  $b_{II}~~$ & $DM$~~~~~ & Distance & $\alpha$~~ & $\beta$~~ & $\log C_{n_e}^2 $\\
      & (deg) & (deg) & (pc cm$^{-3}$) & (kpc)~~~ & & & \\ \hline \hline
B1557$-$50      & 330.69  & 1.63       &    260.56   &    5.66  & $3.87\pm0.09$ & {\it *4.15$\pm$0.2 } & $-1.76$ \\ 
 B1641$-$45     & 339.19  & $-$0.19    &     478.8   &    6.09  & 3.81          &  {\it *4.18}  & $-0.14$ \\ 
 J1740+1000     & 34.01   & 20.27      &     23.85   &    1.36  & 4.59$\pm0.8$  &  3.5$\pm1.2$  & $-1.11$\\ 
 B1740$-$31     & 357.30  & $-$1.15    &    193.05   &    3.65  & 4.49$\pm0.38$  &  3.61$\pm0.63$  & $-1.78$\\ 
 B1750$-$24     & 4.27    & 0.51       &       672   &   10.18  & 4.06$\pm0.77$  &  3.9$\pm1.5$  & $-0.57$\\ 
J1751$-$3323    & 356.83  & $-$3.38    &     296.7   &     9.27 & 4.36          &  3.7 & $-1.19$\\ 
 B1758$-$23     & 6.84    & $-$0.07    &    1073.9   &   13.49  & 4.92$\pm0.11$         &  3.4$\pm0.15$ & $-1.05$\\ 
 B1800$-$21     & 8.40    & 0.15       &    233.99   &    3.94  & {\it 3.25}     &  {\it *5.67}  &$-1.96$ \\ 
 J1809$-$1917   & 11.09   & 0.08       &     197.1   &    3.71  & {\it 2.06}    &  {\it *68}  & $-1.42$\\ 
 B1815$-$14     & 16.41   & 0.61       &     622.0   &    8.10  & 3.97$\pm0.12$ &  4.04$\pm$0.12 & $-1.27$ \\ 
 B1820$-$14     & 17.25   & $-$0.18    &     651.1   &    7.77  & 3.96$\pm0.18$ &  4.03$\pm$0.37  & $-1.88$\\ 
 B1822$-$14     & 16.81   & $-$1.00    &       357   &    5.45  & 3.77$\pm0.24$ &  4.25$\pm$0.54  & $-1.25$\\ 
 B1823$-$11     & 19.80   & 0.29       &    320.58   &    4.83  & 3.02           &  {\it *5.92}  & $-1.48$\\ 
  B1823$-$13    &  18.00  &  $-$0.69   &      231.0  &     4.12 &  {\it1.41$\pm$0.7}  &{\it *$-$4.8$\pm$4.6}  & $-1.65$\\ 
 B1828$-$11     & 20.81   & $-$0.48    &    161.50   &    3.58  & 3.17$\pm0.26$ &  {\it *5.41$\pm$0.88*}  & $-1.42$\\ 
  B1830$-$08    &  23.39  &  0.06      &        411  &     5.67 & {\it 2.13$\pm$0.32}          & {\it*33$\pm$10}  & $-1.65$\\ 
 B1832$-$06     & 25.09   & 0.55       &     472.9   &    6.44  & 4.37$\pm0.30$           &  3.68$\pm0.5$   & $-0.84$\\ 
  B1834$-$04    &  27.17  &  1.13      &      231.5  &     4.62 &  {\it 9.76}         & {\it 2.5}  & $-1.54$\\ 
 J1835$-$1020   &  21.98  &  $-$1.30   &      113.7  &     2.57 &  {\it 0.20$\pm$0.05} & {\it *$-$0.22$\pm$0.11} & $-2.01$\\ 
 B1838$-$04     & 27.82   & 0.28       &   325.487   &    5.17  & 3.91$\pm0.14$          & 4.09$\pm$0.29  &$-1.33$\\
J1842$-$0905    & 23.81   & $-$2.14    &     343.3   &     7.41 & {\it 5.63}          &  {\it 3.10} & $-2.69$\\
 J1857+0143     & 35.17   & $-$0.57    &      249    &    5.18  & 4.69          &  3.4  &$-0.43$ \\
J1901+0510      & 38.74   &  0.03      &     429     &     8.47 & 4.14          &  3.87  &$-2.54$ \\
 J1907+0918     & 43.02   &  0.73      &     357.9   &     7.68 & 3.64$\pm0.8$ &   4.4$\pm$1.9  &$-2.31$ \\
J1910+0728      & 41.74   &  $-$0.77   &     283.7   &     6.04 & 2.77          &  *7.19  & $-2.02$\\ \hline
\end{tabular}
\end{table*}

Using the data from Table~\ref{tau_table}, and assuming that the values of the measured scatter times $\tau_d$ are indeed governed by a power-law, we performed a weighted least-squares fits to the observational data to find the actual spectral indices for he pulsars from our sample. To ensure the best results and proper uncertainty analysis we actually fitted a power-law function ($\tau_d = 10^{\, -\alpha \log\nu+C}$) to our measurements. The results of our fits - presented in a log-log scale for better clarity - are shown in Fig.~\ref{plot_tau_vs_freq} as the solid lines. The dashed lines on the plots represent the expected slope of a thin screen Kolmogorov-like dependence. One can clearly see deviations from the expected power law, which - as it was mentioned in the Introduction - happens quite often for high dispersion measure pulsars.

Table~\ref{table_alpha} presents the results of our fits for the spectral index $\alpha$ for pulsars from our sample, along with some of the basic parameters - their galactic coordinates, the dispersion measure and a distance estimate - from the \citet{taylor93} galactic electron density model. Apart from the scatter time spectral index $\alpha$ the table includes also the spatial electron density spectral index $\beta$ ($\beta = 2 \alpha/[\alpha-2]$), and the estimated fluctuation strength for the pulsar's line-of-sight $C^2_{n_e}$, which we calculated the same way as L04. Note that the
mentioned relation between $\alpha$ and $\beta$ is valid only for $\alpha \geq 4.0$. The un-physical values of $\beta$ are indicated by an asterisk in the table; we left the values un-marked when the value of $\alpha$ is greater  or equal to 4.0 within the error estimates.

The table also includes the $\alpha$ and $\beta$ values calculated for the pulsars for which we had the measurements of the pulse broadening $\tau_d$ available (or usable) at two observing frequencies only - these are not shown in Figure~\ref{plot_tau_vs_freq}, and quoted as error-less in the table. We however decided to use the values obtained this way to improve our statistics.

In a few cases to achieve the final fit for the spectral index $\alpha$ we omitted individual measurements - these are indicated in Table~\ref{tau_table}, and can be clearly identified in Figure~\ref{plot_tau_vs_freq}. Usually (but not in all cases) these were the highest observing frequency values, which may have suffered from various flaws of the measurement method that was used (see the discussion in Section~\ref{errors}). In other cases (like the second-highest frequency values in B1758$-$23 and B1832$-$06) inclusion of the omitted points results significant reduction of the quality of the fit, and additionally yields much higher slopes, which we consider unrealistic.

As one can see for some of the pulsars the values of the scattering spectral index $\alpha$ that resulted from our fits are definitely not the real values which range from $\alpha = 0.20$ for PSR~J1835$-$1020 (the profile evolution of this pulsar shown in Fig.~\ref{1835_prof}) to $\alpha = 9.76$ for PSR~B1834$-$04. This is indeed very far from the expected values of $\alpha \approx 4.4$, and results in erroneous, in some cases even un-physical values of the spatial density index $\beta$ (i.e. negative values of $\beta$). Clearly, those pulsars suffer from inaccurate scatter time measurements, which can happen in our method for various reasons, as it was mentioned before in Section~\ref{errors}. In case of PSR~J1835$-$1020 the reason behind the erroneous value may be the fact that over the whole range of frequencies, from 600~MHz to 8.35~GHz, there is only a very small change in the profile shape and only the lowest frequency profile shows any hint of a scattering tail at a first glance. This may have been caused by a heavy dispersion smearing in our data, but for this pulsar $DM = 113$~pc~cm$^{-3}$ and since the period of the pulsar is 302~ms one would expect the dispersion to have a minimal effect. On the other hand a low value of the dispersion measure translates (usually) to small pulse scatter-broadening. The effect of the dispersion smearing, however small, may be comparable to the scatter broadening, which results in un-real values of the scatter time $\tau_d$ when applying our measurement method. 

The other pulsar mentioned above, PSR~B1834$-$04 suffers from having its pulse-broadening measured at only two frequencies: 610~MHz - where the scattering is strong (conf. profile in the Appendix), and 1.1~GHz where the scattering tail barely exists. Closer inspection of the 1.1~GHz profile suggests that it is a multi-component case with the leading component weaker than the trailing one and both quite broad - either intrinsically or due to the dispersion-broadening. We tried a few different approaches to this profile, to be able to estimate the scatter time. Since we did not have any higher frequency profiles we could not attempt to model this profile as a two-gaussian-component - attempts of modeling with the distance between components being a free parameter of the model have failed. We also tried to limit the profile-longitude range of the fit, allowing the fit to be performed only for the supposed second component (i.e. performed the fit for the data points starting just prior to the profile maximum). In this case the gaussian component of the convolution apparently dominated the fit - since the profile did not have the left part at all - the gaussian component parameters were found by fitting to the right-side, scattered part of the profile, leaving very little to the apparently small scattering effect. This resulted probably in an underestimation of the scatter time, which in turn led to a very high slope in the scattering spectrum.

We excluded the values obtained for these pulsars from further analysis, along with the results obtained for 4 other pulsars. In case of PSR~B1823$-$13 ($\alpha = 1.41$) we had a very narrow range of frequencies available (0.9 to 1.6~GHz, conf. Fig.~\ref{plot_tau_vs_freq}). For B1800$-$21 we got an ``acceptable'' result for $\alpha = 3.25$ when omitting the lowest frequency measurement (very poor profile - see Appendix), but both the 925~MHz and 1.4~GHz EPN profiles on which this measurement would be based show very little scattering and are severely smeared (low phase resolution). Additionally our fits to 2.6~GHz data profile (shown in the Appendix) did not yield any reliable scatter time value. Another case, PSR~J1842$-$0905, had $\tau_d$ measured at only two frequencies. With a multi-component profile it is susceptible to erroneous measurements when using our simplified method, hence we decided to exclude the resulting (and in fact very high) value of $\alpha = 5.63$ from further analysis.

The last excluded pulsar, PSR~B1830$-$08 ($\alpha = 2.14$), was a case that showed no apparent scattering at 4.85~GHz (hence the $\tau_d$ value may be over-estimated, see discussion in Section~\ref{errors}). We also used our observation at 610~MHz, and a EPN profile from 1.6~GHz which was severely broadened, probably due to dispersion smearing/low phase-resolution. Additionally it is probably a multi-component profile with a weak component trailing the profile maximum. Such an unidentified component can also result on an overestimation of the scatter broadening when using our method of analysis. With scatter time values overestimated at both 1.6~GHz and 4.8~GHz, this resulted in a very low spectral slope and a low value of the spectral index $\alpha$.

\begin{figure}
\resizebox{\hsize}{!}{\includegraphics{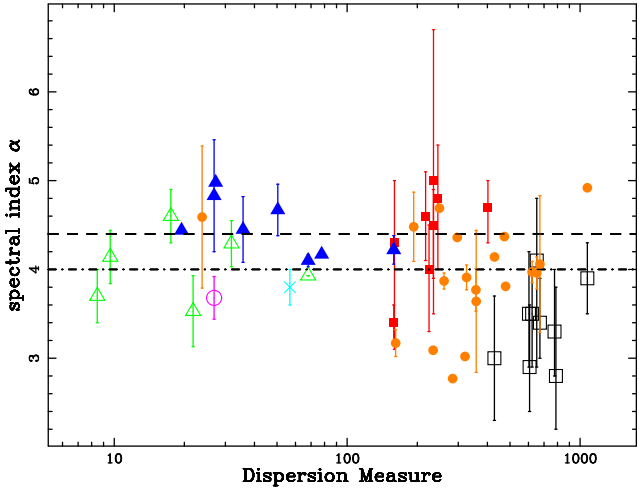}}
\caption{A plot of the spectral index of pulse broadening $\alpha$ versus the dispersion measure. Empty squares (black) - L01, filled squares (red) - L04, empty triangles (green) - \citet{johnston98}, filled triangles (blue) - \citet{cordes85}, cross (light blue) - \citet{kuzmin02}, empty circle (purple) - \citet{lewan11}, filled circles (orange) - this paper.\label{alpha_dm}}
\end{figure}

\begin{figure}
\resizebox{\hsize}{!}{\includegraphics{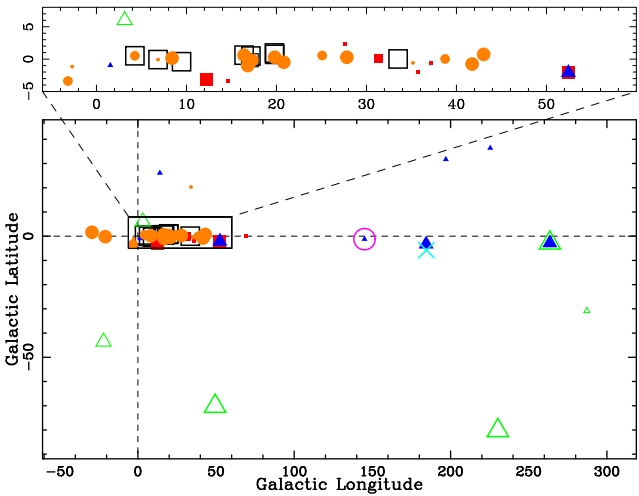}}
\caption{Pulsars with measured scattering spectral index as seen on the sky in the Galactic coordinates. 
Symbol and colour coding same as on Fig.~4. The size of the symbol denotes the value of 
$\alpha$: smallest - $\alpha>4.4$; medium $4.0<\alpha<4.4$, largest $\alpha<4.0$. 
The top sub-panel shows a close-up of the Galactic Center region.\label{alpha_lb}}
\end{figure}

The remaining pulsars were used for further analysis. A few of the pulsars from our sample were previously observed by other projects, namely L01. Our results differ somewhat from L01 data. For PSR~B1750$-$24 our $\alpha = 4.06\pm0.77$ while for L01 $\alpha_{\rm L01}=3.4$. We used the profiles obtained at exactly the same frequencies (1.4~GHz, 1.6~GHz and 2.6~GHz), the scatter-broadening measurements are however quite different - for example at 1.4~GHz $\tau_d$ is 49~ms for L01, versus  our value of 85~ms, and at 1.6~GHz it is 25~ms vs our measurement (based on the profile from the EPN database) of 43~ms. For the 2.6~GHz we used the profile from our Effelsberg observations, and got a value of $\tau_d = 13$ms, different from the one quoted by L01 (5.2~ms), but looking at our profile (see Appendix) which is showing a significant scattering,  we are quite convinced that our measurement is reliable, although due to the poor quality of the data the error estimate is of the order of 30\%. This error estimate in turn makes the actual fit to almost completely omit that point, and results in a big error estimate in our $\alpha$. What is most puzzling is that the our measurements of $\tau_d$ differ from the ones provided by L01 by a similar factor at all frequencies, which may be an indication of the inhomogenities in the interstellar medium which may affect the scattering parameters - this was observed for low-$DM$ pulsars by
\citet[][see also references therein]{brisken10}, and while B1750$-$24 is rather a high-$DM$ pulsar ($DM = 672$~pc~cm$^{-3}$) the possibility still exists.

For B1758$-$23 ($\alpha= 4.92\pm0.11$ vs $\alpha_{\rm L01} = 3.9$) the results at 1.4 and 1.6~GHz are quite similar (conf. L01 Table 1 with our Table~1), however we measured much smaller value of scatter broadening at 2.6~GHz: $\tau_d = 0.75$~ms, versus L01's 8.6~ms. We also obtained profile at 4.8~GHz but it shows next to no scattering. For the $\alpha$ fit we decided to exclude the suspiciously low value at 2.6~GHz, which was definitely deviating  from the general trend - its inclusion rises the $\alpha$ even more, but degrades the quality of the fit. The reason behind the discrepancies at higher frequencies may be again the profile asymmetry and rather poor quality of our high frequency measurements. Excluding both high frequency measurements yields almost exactly a Kolmogorov's slope ($\alpha = 4.33$) but the reliability of such result would be very low, as it comes from just two very similar observing frequencies (1.4 and 1.6~GHz).

The last pulsar common for our project and L01 is B1815$-$14 ($\alpha = 3.94\pm0.12$ vs $\alpha_{\rm L01} = 3.5$). In this case we believe in our value, as L01 did not have the measurement at 0.6~GHz, and for the other frequencies (1.0, 1.4 and 1.6~GHz) our data agrees with L01 within error estimates.

Combining our results with the data found in the literature we decided to recreate the plot of L01 and L04 showing $\alpha$ vs. $\log(DM)$, and the result is shown in Figure~\ref{alpha_dm} (for the references see the Figure caption). This plot includes the results of the scatter time spectral index obtained from direct scattering measurements, as well as the values of $\alpha$ obtained from scintillation observations: according to the theory of scintillation \citep[see for example ][]{rickett90} the decorrelation bandwidth spectral evolution is governed by the same spectral index $\alpha$.

After adding our data points to the picture, we can confirm the general idea of L01 and L04 that for high-dispersion measure pulsars the spectral behaviour of the scatter time spectral index starts to deviate from the Kolmogorov's theory predictions. Our data added a significant ``scatter'' to the picture for mid-to high $DM$ pulsars, however a good number of them is showing a clear departure from the expected value of  $\alpha=4.4$. It seems also, that a significant number of data point starts to deviate from the theory predictions starting with $DM \approx 230 \div 250$~pc~cm$^{-3}$, and not as L01 and L04 suggested for $DM >300$~pc~cm$^{-3}$. It also seems, that least a few of our sources still have their values of $\alpha$ in a range between 4.0 and 4.4, i.e. in quite  good agreement with the Kolmogorov's theory predictions. This is in a contrary to L01 and L04, which had only one point in a $DM > 300$ range with a value of $\alpha \geq 4.4$. As for our points that lie far above the Kolmogorov's line, one of them is the afore mentioned PSR~B1758$-$23 ($DM = 1073.9, \alpha=4.92$), which may have suffered from erroneous scatter time measurements, and PSR~J1842$-$0905 ($DM = 343.3, \alpha=5.65$) - a pulsar with measurement at only two observing frequencies (600~MHz and 1.4~GHz), which has a multi-component profile, and apparently shows a significant scattering only at the lower frequency (conf. profiles in the Appendix). We believe that the data points belonging to those two pulsars may be disregarded in further interpretation of the figure.

\begin{figure*}
\resizebox{0.9\hsize}{!}{\includegraphics{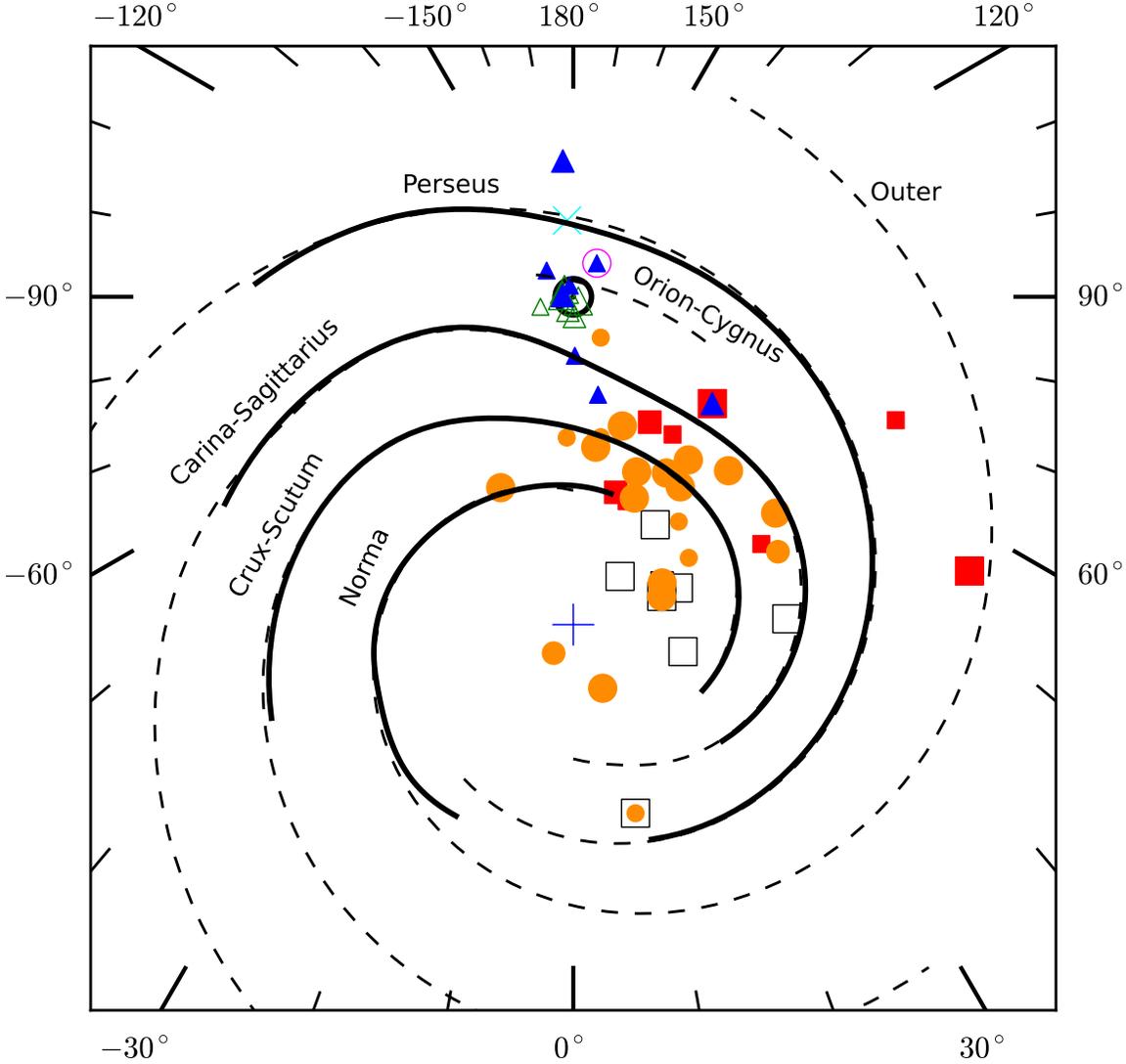}}
\vskip-3mm
\caption{Pulsars with measured scattering spectral index in the galactic X-Y coordinates. Symbol, colour and size coding same as on Fig.~4 and 5.\label{alpha_xy}}
\end{figure*}

Following L04 we decided to plot our pulsars against the galactic coordinates as well, and the plot is shown in Figure~\ref{alpha_lb}. To retain the information about the value of the scattering spectral index we used three sizes of graph markers. The smallest ones denote pulsars with $\alpha > 4.4$, the medium-sized markers were used for pulsars with $\alpha$ between 4.0 and 4.4. Both those groups, after the exclusion of pulsars with unrealistically high values of $\alpha$, can be considered to consist of sources with ``relatively good agreement'' with the Kolmogorov's theory predictions. Finally, the largest size graph markers were used to emphasize pulsars with values of the scatter time spectral index lower than 4.0, i.e. the most interesting cases that clearly deviate from the commonly used theory. The sub-panel in the figure shows a close-up of the galactic center region.

The $\alpha$ versus galactic coordinates plot however does not show any significant trends, since it can not represent the 3rd dimension, the distance from the observer. Therefore we decided to plot our data against the pulsar position in the Milky way Galaxy, using the values of the $X$ and $Y$ coordinates that can be found in the ATNF Pulsar Catalogue\footnote{Available at {\tt http://www.atnf.csiro.au/people/pulsar/psrcat/}, \citet{manchester05}} (for all the mid- and high-$DM$ pulsars these were inferred from the $DM$-distance estimate). The result is shown in Figure~\ref{alpha_xy}, and the graph-marker size coding used is the same as in Figure~\ref{alpha_lb}, i.e. the largest graph markers represent the objects with values of $\alpha$ much lower than the theoretically predicted value of $4.4$. The points are plotted against the spiral arm structure taken from the \citet{taylor93} model.

\section{Discussion and Conclusion}

As it was shown above, for some of the pulsars the observed frequency dependence of the scatter time does not comply with the predicted spectral index of $4.4$. As it was shown in Fig.~\ref{alpha_dm}, the really significant deviations start for pulsars with $DM > 250$~pc~cm$^{-3}$, i.e. the most distant pulsars that had their scattering parameters measured. However some of the close-by pulsars also show a significant departure from the predictions. Clearly there can be no single explanation for both of these cases, and the reasons for the deviations may be different. Below we will discuss some of the possibilities.

One has to remember, that the prediction of the $\alpha = 4.4$ is valid only for a model, which uses: 
\begin{description}
\item{{\it (i)\/} a single thin screen distribution of the ISM, or a thick medium, uniformly distributed between the pulsar and the observer; }
\item{{\it (ii)\/} the Kolmogorov's spatial electron density spectrum (with $\beta = -11/3$); }
\item{{\it (iii)\/} an assumption that the {\it inner scale} ($k_i^{-1}$, see Eq.~\ref{elec_dens_full})  of 
the density fluctuations is negligibly small, and the {\it outer scale} ($k_o^{-1}$) is large enough to not affect the scattering spectrum.  }
\item{{\it (iv)\/} and the assumption that the ISM turbulence is isotropic and homogeneous.}
\end{description}

If even a single of these assumptions is not fulfilled, 
the expected $\tau_d$ spectral index may change.

Probably the simplest explanation of the high-$DM$ pulsars non-compliance with the simple model would be the fact that for the distant objects the assumption {\it (i)} of a single thin screen is unrealistic, and neither the uniformly distributed ISM (i.e. the thick screen) seems valid. This explanation was already invoked by L01 and L04. Fig.~\ref{alpha_xy} clearly shows that for most of the distant pulsars which show the deviations from the simple model, the line-of-sight crosses (or includes at least a part of) two or more of the Milky Way spiral arms. Obviously, the probability of encountering a high-electron density cloud/region (like an H~II region) is significantly greater within a spiral arm than in the space between them; therefore we believe that for most of these pulsars the line-of-sight would cross more than one high-electron density region and the radiation would undergo multiple scatterings. No reliable model was developed for multiple scattering screens case as of yet, this matter clearly requires further investigation.  One can expect that for the multiple screens along the line-of-sight we will also get predictions different from the simple model. It is easy to understand that while the scattering on the first screen that the pulsar radiation encounters may comply with the simple model, the second screen (and any additional screen after that) would interact with rays already scattered by the first screen, which won't have the same ``symmetry'' as the original pulsar radiation. Also, the lower frequency radiation will be scattered more by the first screen, compared to the higher frequencies, which may influence the way it will interact in the second screen. Clearly, the multiple scattering screens case requires further study, and proper modelling, which we intend to do in the near future.

Another explanation of the discrepancy between the basic model and the measured $\tau_d$ spectral index may be the spectrum of the spatial density fluctuations {\it (ii)} is actually not the Kolmogorov's spectrum, i.e. $\beta$ is different from $11/3$. This was thoroughly discussed for the purpose of the interstellar scintillation observations analysis \citep[for a brief summary see ][]{gupta00}, where the spectral index of the electron density distribution affects several observational parameters. No conclusions can be drawn however, most of the observations indicate that the Kolmogorov's theory is working, however there are also a few deviations - see for example the recent results of \citet{lewan11} for PSR~B0329+54. The possibility and effects of other spatial electron density distributions, and their predictions for the $\tau_d$ spectral index - which is the same as the decorrelation frequency spectral index in the scintillation analysis - was discussed by \citet{romani86}. For realistic models one can get the slope of the $\tau_d$ frequency dependence between $4.0$ and $4.7$. One has to note however that for the majority of low-to-mid $DM$ pulsars (for these one can reasonably assume that a single scattering screen model will be an accurate approximation) the Kolmogorov's model seems to work properly, which is also somewhat visible in our Fig.~\ref{alpha_dm}, for pulsars with $DM$ between 30 and 250~pc~cm$^{-3}$.

There is of course the issue of some of the closest pulsars, with $DM$ lower than 30~pc~cm$^{-3}$, and the fact that a significant number of them shows a clear deviation from the simple model with Kolmogorov's spectrum. But since for the pulsars a bit further away using the Kolmogorov's spatial electron density spectrum seems to be working properly, there is no reason to believe that it won't work for close pulsars. The reasons for those discrepancies is rather the peculiar distribution of the electron matter along certain  lines-of-sight, caused by the Local Interstellar Medium \citep[as it was discussed by ][]{bhat99}, i.e. rather a non-compliance with the first assumption of the simple model {\it (i)}, than the second {\it (ii)}. 

The next source of discrepancy may be an incorrect assumption that {\it (iii)} the inner scale of the electron density fluctuation spectrum is negligibly small compared to the actual fluctuation size, which (along with the assumption that the outer scale is very large) simplifies the general spatial density spectrum (see Eq.~\ref{elec_dens_full}) to a power-law. As we mentioned in the Introduction (following \citealt{rickett09}), the outer scale should be less of an issue, as the hypothesis of a steeper spectrum was proven incorrect. Approaching the outer scale causes the spectra of spatial electron density to be less steep, which in turn would cause the scatter time spectrum to steepen above $\alpha=4.4$. For the pulsars from our sample, after excluding the sources that clearly suffer from erroneous measurements, only PSR~B1758$-$23 shows the slope significantly steeper than the Kolmogorov's spectrum (i.e. $\alpha=4.92$); but even in this case it is mostly caused by two suspicious high-frequency measurements - excluding them causes $\alpha$ to drop to a value of 4.33.    

On the other hand it is possible that the inner scale ($k_i^{-1}$) is not small enough. This would cause the electron density spectrum to steepen when approaching this value, and as \citet{romani86} showed, the steeper density spectrum yields less steep slopes for the scatter time frequency dependence. As we mentioned in the Introduction, for an individual source this may cause the spectral index $\alpha$ to be Kolmogorov-like at high frequencies, while drooping to 4.0 at low frequencies. In our sample we found only four sources, for which we feel safe to try to check for this effect, i.e. the sources that had scatter time measured at four or more frequencies. The results are shown in Table~\ref{two_alpha}.

\begin{table}
\caption{Low- and high-frequency $\tau_d$ spectral slopes. The error-less values were estimated from just two data points.\label{two_alpha}}
\begin{center}
\begin{tabular}{lll}
\hline
Pulsar & $\alpha\ (\nu \leq 1060)$ & $\alpha\ (\nu \geq 1060)$ \\
\hline
\hline
B1557$-$50 & $3.85\pm0.11$ & $3.96\pm0.47$ \\
B1815$-$14 & 4.29 & $3.71\pm0.02$ \\
B1822$-$14 & 4.07 & $3.19\pm0.02$ \\
B1832$-$06$^{\star}$ & 4.04 & $4.34\pm0.04$ \\ \hline
\multicolumn{3}{l}{$^{\star}$ the actual division used was at 1.4~GHz}
\end{tabular}
\end{center}
\end{table}

As one can see only for PSR~B1832$-$06 we got the desired result. In case of PSR~B1815$-$14 and PSR~B1822$-$14 the spectra seems to rather steepen at lower frequencies, but this may be due to a fact that for both of these pulsars the low-frequency part of the spectrum consisted of only 2 points, and the actual error estimate at the lowest frequency is rather large in both cases (ca. 30\%). For PSR~B1557$-$50 the spectral index seems to be consistent over the entire range of observing frequencies, and in fact very close to 4.0. Overall we feel that this analysis is inconclusive. Our data does not ensure a satisfactory frequency coverage at the lower end of the spectrum, and additionally that would not be easy to achieve. At the frequencies lower than 610~MHz these pulsars would be not observable by standard means due to very high scattering, and also there are not many telescopes with receivers that would cover range between 610~MHz and 1.4~GHz at the frequencies other than the ones we used. The high frequency observations on the other hand suffer from the fact  that our observing project was not designed with the scattering phenomenon  in mind  and the quality of the measurements definately can be improved by performing a dedicated observing project (which we consider for the future).

Finally, the discrepancies between the ``expected'' spectral slope of 4.4 and the measurements may be due to the fact, that the theory (presented in the Introduction) assumes that {\it (iv)\/} the turbulence in the ISM is isotropic and homogeneous. It was shown however that in case of the solar wind plasma its spatial anisotropy can cause the turbulence spectrum to become flatter than $\beta=11/3$ \citep{harmon05}, which is impossible to get using homogeneous isotropic turbulence. In case of the ISM theory it was shown that the transverse structure of the screen will affect the observed spectrum. The configuration of even a single screen can affect the predicted value of the $\tau_d$ spectral index. For example, for a  ``truncated'' screen one can get an expected index of $\alpha$ of $4.0$ or less \citep{cordes01}. This is because the simple model assumes that the screen is infinite in the direction transverse to the line of sight, which may not be the case in the real astronomical objects (like for example scattering by the narrow filaments inside a supernova remnant). In the simple model with transversely-infinite screen one can expect a perfect symmetry. If the screen ends abruptly in the transverse direction, one may expect the high frequency radiation (which is less scattered) to behave like it would encounter an infinite screen. However for the low frequency radiation, some part of it that would be scattered towards us by an infinite screen, will not reach us because it won't encounter the scattering screen at all. This obviously will alter  both the shape of the profile (and therefore the measured $\tau_d$), as well as the amount of radiation that we get, possibly showing an increasing deficit of radiation the lower the observing frequency is.

The ISM turbulence may be not only anisotropic, but also inhomogeneous. If the turbulence is inhomogeneous the scattering parameters will change with time. In fact they can change on the time it takes the line of sight from the Earth to the
pulsar to move through the scattering disc. This might be a month or so for a typical pulsar and can be estimated given the distance and proper motion of the pulsar. Therefore it would be most beneficial to perform the multi-frequency observations in at least quasi-simultaneous way; sadly it was not the case for our data (see the end of Section~\ref{obs}). This of course can easily falsify the $\alpha$ measurements, and may cause the spectral index $\alpha$ to drop below 4.0, or rise above 4.7, i.e. out of the bounds predicted by the homogeneous turbulence theory \citep{romani86}. At least in one case - PSR~B1750$-$24 - we can claim the existence of the inhomogeneity, as our results significantly differ from the ones shown by L01 (and for almost all the frequencies by the same factor); also this resulted in different values of estimated spectral slope - L01's result of $\alpha=3.4$ is way below any homogeneous theory predictions while we got $\alpha=4.06$.

Overall, one cannot exclude the possibility that all four {\it (i - iv)\/} assumptions may be wrong (to a degree), however we believe that the biggest impact on the observed scattering properties comes from the screen effects and the supposed inhomogeneity of the turbulence (which may directly falsify the measured scatter time spectral slope). Luckily, the latter can be at least partially excluded by performing quasi-simultaneous multi-frequency observations. The distribution of the electron matter along the lines-of-sight however will always play the most important role. For the most distant pulsars the easiest explanation would be the existence of multiple screens along their lines-of-sight (which in case of some of our pulsars is highly probable - conf. Fig.~\ref{alpha_xy}). The model of multiple-screens scattering needs to be developed to attempt to confirm that it would indeed ``flatten'' the spectra of the  $\tau_d$ frequency dependence, especially in cases when it drops below $\alpha=4.0$ in the observations.

One has to also note that whatever is the reason for the deviation of the $\tau_d$ slope from the value of $\alpha = 4.4$, those deviations are always making the dependence less steep (down to 3.0), meaning that at the lower observing frequencies the scattering effect will be smaller than the one predicted by the single-screen Kolmogorov's model. This fact may be really important for the future pulsar search surveys, as it would  make the most distant pulsars (strongly affected by scattering) actually easier to detect. That would in turn allow to lower the observing frequency of the surveys (i.e. search for pulsars at the frequencies where they are stronger), without risking that the pulsars will be completely ``smeared'' by the scattering phenomenon.

\section*{Acknowledgments}
This paper was supported by the grant DEC-2012/05/B/ST9/03924 of the Polish National Science Centre. MD is a scholar within Sub-measure 8.2.2 Regional Innovation Strategies, Measure 8.2 Transfer of knowledge, Priority VIII Regional human resources for the economy Human Capital Operational Programme co-financed by European Social Fund and state budget. We thank Jim Cordes for his remarks about this paper, and the software used to create one of the figures. We are also extremely grateful to the anonymous referee for extensive remarks and explanations of the scattering theory, that helped to greatly improve the completness of the discussion and the overall quality of this publication.

\appendix
\section*{Appendix: Pulsar Profiles and additional data}

In this appendix we present the results of the flux density and the pulse width measurements for the 
pulsars that were observed over the course of our projects. The measurements are summarized in Table~\ref{app_table}.
These values were never previously published.

One has to remember however, that the quoted pulse widths may be affected by both the interstellar scattering 
phenomenon (especially at lower frequencies), and the dispersion smearing which is due to very wide bandwidths 
used at the higher observing frequencies (2.6~GHz and above) - see main body of the paper for explanation.

We also present the pulse profiles we obtained during our observing campaign, 
which were used for the purposes of performing the scatter time measurements (along with the 
profiles obtained from the EPN and ATNF databases).

\setcounter{table}{3}

\begin{table}
\caption{Flux density measurements, pulse widths and the profile classification for observed pulsars\label{app_table}}
\begin{tabular}{lccccccc}\hline
      & Class. & Freq. & $S$ & $W_{10}$ &  $W_{50}$\\
Pulsar    & &(MHz) & (mJy) & (deg) & (deg)  \\ \hline \hline
B1557$-$50   &  St$^1$   & 325    &  $128\pm38$    &  283   & 125  \\
             &           & 610    &  $66\pm13$     &  198   &  71  \\
             &           & 1060   &  $26.6\pm1.4$  &  39    & 11   \\
 B1641$-$45  &  St$^1$   & 325    & $630\pm53$     &  155      &  62    \\
             &       &   1060     & $323\pm52$     &  23    &  10  \\
 J1740+1000    & 3C$^2$& 610    & $6.1\pm2.7$    &  64    & 28   \\
               &       & 1170   & $0.9\pm0.3$    &  40    & 25   \\
 B1740$-$31    &  1C   &  325   & $7.6\pm0.6$    &  92    & 28   \\
               &       &  610   &$10.6\pm2.6$    &  22    &  8   \\
               &       &  1060  & $3.0\pm0.3$    &  12    &  7   \\
 B1750$-$24    &2C$^3$ &  2640  &  $1.10\pm0.02$ &  39    & 17     \\
 B1758$-$23    &2C$^3$ &  2640  & $1.70\pm0.03$  &  53      &  32    \\
 B1800$-$21    & 2C    &  610   & $18.2\pm0.9$   &  204   & 82   \\
               &       &  2640  & $7.0\pm0.1$    &  109   & 27   \\
J1806-2125     & 1C$^4$&  1170  & $1.5\pm0.5$    &  190   & 250  \\
 J1809$-$1917  & 2C$^4$&  1170  & $1.1\pm0.3$    &  110   & 72   \\
               &       &  2640  & $2.3\pm0.3$    &  106   & 34   \\
               &       &  4850  & $0.9\pm0.1$    &  96    & 68   \\
 B1815$-$14    &1C$^3$ &  610   & $6.7\pm1.9$    &  300   & 139  \\
               &       &  1060  & $8.6\pm0.2$    &  165   & 41   \\
 B1820$-$14    & 1C$^6$&  610   & $8.8\pm0.6$    &  246   & 48   \\
               &       & 1060   & $1.2\pm0.1$    &   58   & 13   \\
               &       & 2640   & $0.30\pm0.03$  &   49   & 32   \\
 B1822$-$14    &1C$^3$ & 610    & $1.8\pm0.3$    &  145   & 52   \\
               &       & 1060   & $2.4\pm0.5$    &  62    & 22   \\
 B1823$-$11    &2C$^6$ & 325    &                &  102   & 32   \\
  B1823$-$13   &2C     & 1060   & $3.6\pm0.8$    &  155   & 38,34\\
               &       & 2640   & $3.6\pm0.1$    &  121   & 32,32\\
 B1828$-$11    & 1C    & 325    &                &  306   & 104  \\
               &       & 610    & $1.5\pm0.2$    &  25    &  6      \\
               &       & 1060   &                &  10    &  4      \\
               &       & 2640   & $0.4\pm0.1$    &  10    &  5      \\
               &       & 4850   & $0.06\pm0.01$  &  11    &  6      \\
J1828-1101     & 2C$^4$& 2640   & $0.4\pm0.1$    &  124   & 117  \\
               &       & 4850   &$0.06\pm0.01$   &  119   & 95   \\
  B1830$-$08   & 2C$^3$& 610    & $18.5\pm3.2$   &  264   & 97   \\
 B1832$-$06    & 1C$^6$& 1060   &                &  188   & 87   \\
               &       & 2640   & $0.70\pm0.03$  &  32    & 22   \\
               &       & 4850   & $0.04\pm0.01$  &  29    & 21   \\
  B1834$-$04   & 2C   & 610    & $5.8\pm1.3$    & 85     & 22   \\
               &       & 1060   &                & 22     & 9     \\
 B1834$-$06    & 2C    & 1170   &                &  60    & 18   \\
  J1835$-$1020  & 1C   & 610    & $5.8\pm1.7$    & 16     & 8      \\
                &      & 1170   & $2.7\pm0.4$    & 15     & 6   \\
                &      & 2640   & $0.93\pm0.04$  & 14     & 6   \\ 
                &      & 4850   & $0.22\pm0.07$  & 12     & 7   \\ 
                &      & 8350   & $0.19\pm0.01$  & 10     & 4   \\
B1838$-$04    & 1C$^6$ & 610    & $12.7\pm0.5$   & 154    & 40 \\
              &        & 1060   & $2.7\pm0.8$    & 25     & 10  \\
B1849+00      & 3C$^3$ & 1060   &                & 198    & 41  \\
              &        & 2640   & $1.3\pm0.3$    & 42     & 22  \\
 J1857+0143   & 1C     & 1170   & $0.5\pm0.1$    & 60     & 50  \\
              &        & 2640   & $0.19\pm0.06$  & 39     & 18  \\
              &        & 4850   & $0.03\pm0.02$  & 28     & 18  \\
J1905+0616    & 1C     & 610    & $2.1\pm0.6$    & 20     &  4    \\
              &        & 1060   & $0.4\pm0.1$    & 16     &  3 \\
 J1907+0918   & 2C$^5$ & 610    & $1.0\pm0.2$    & 22     & 6     \\
              &        & 1170   & $0.20\pm0.06$  & 12     & 4 \\
              &        & 2640   & $0.04\pm0.02$  & 27     & 9     \\ \hline
\end{tabular}
1C,2C,3C - denotes one-, two-, or three-component profile respectively \\
Profile classification source :$^1$ \citet{rankin83}, $^2$ \citet{mclaughlin02},\\
$^3$ \citet{kijak98},$^4$ PMPS-2 \citep{pmps2} ,$^5$ PMPS-4 \citep{pmps4},$^6$ \citet{gould98}, other - our classification
\end{table}

\setcounter{figure}{6}

\begin{figure}
\resizebox{\hsize}{!}{\includegraphics[angle=-90]{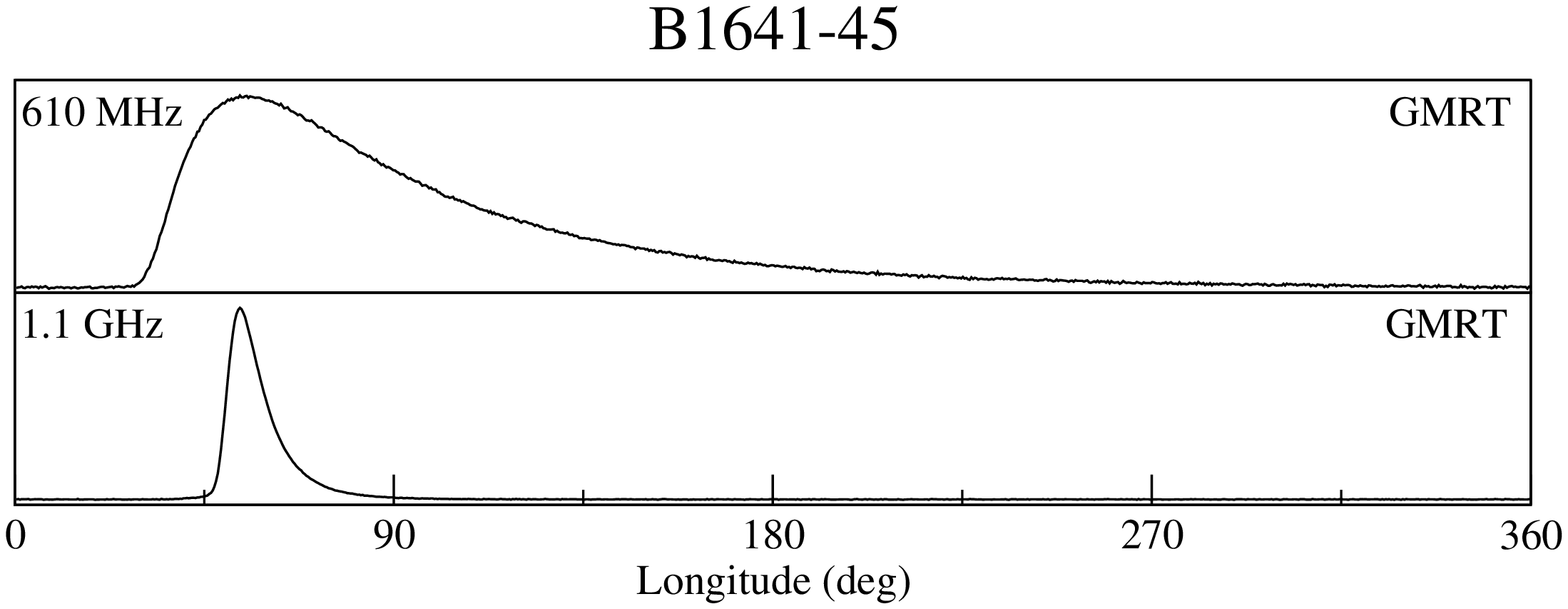}}
\resizebox{\hsize}{!}{\includegraphics[angle=-90]{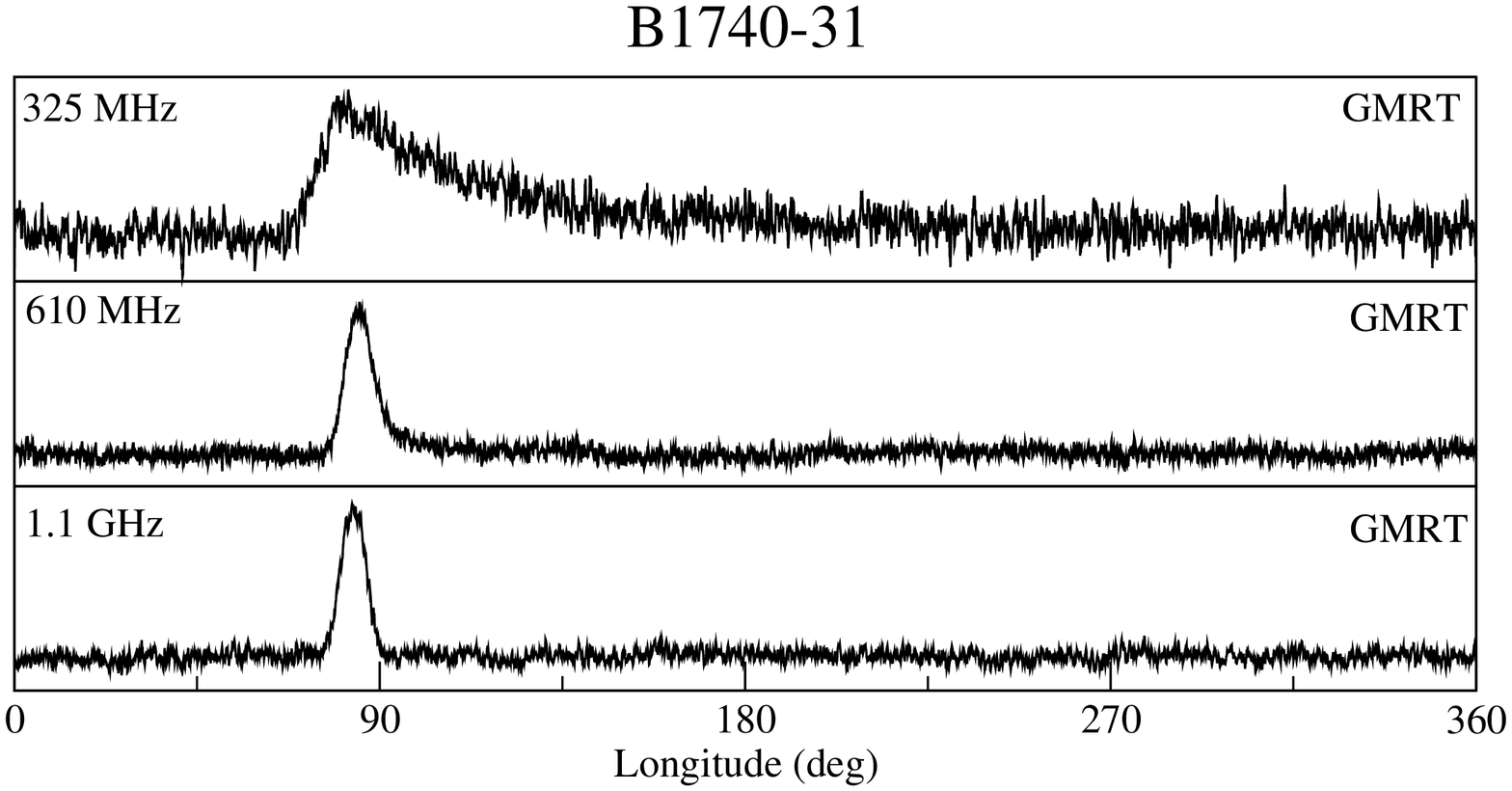}}
\resizebox{\hsize}{!}{\includegraphics[angle=-90]{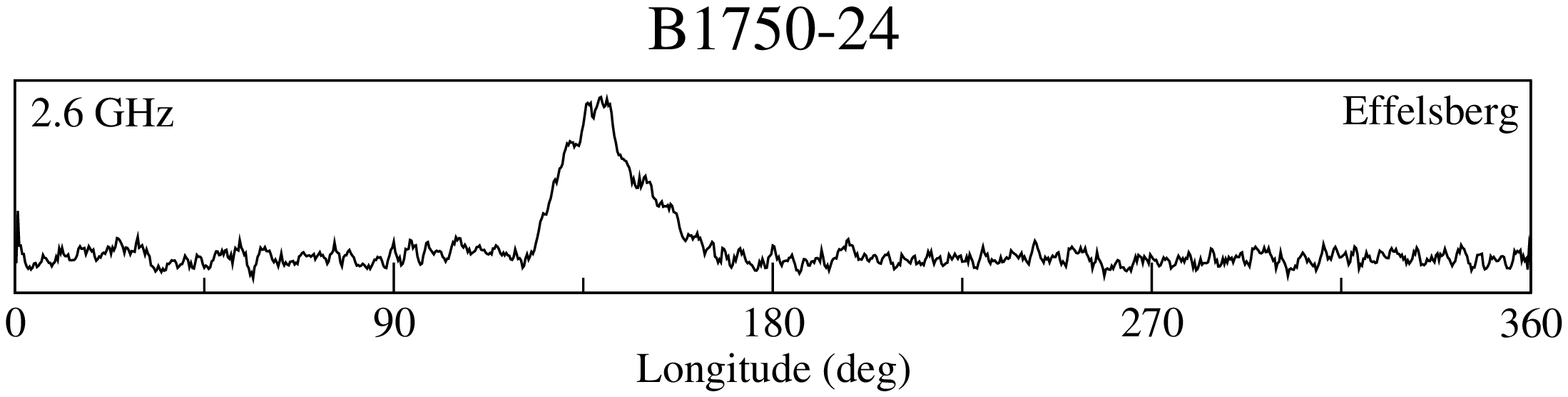}}
\resizebox{\hsize}{!}{\includegraphics[angle=-90]{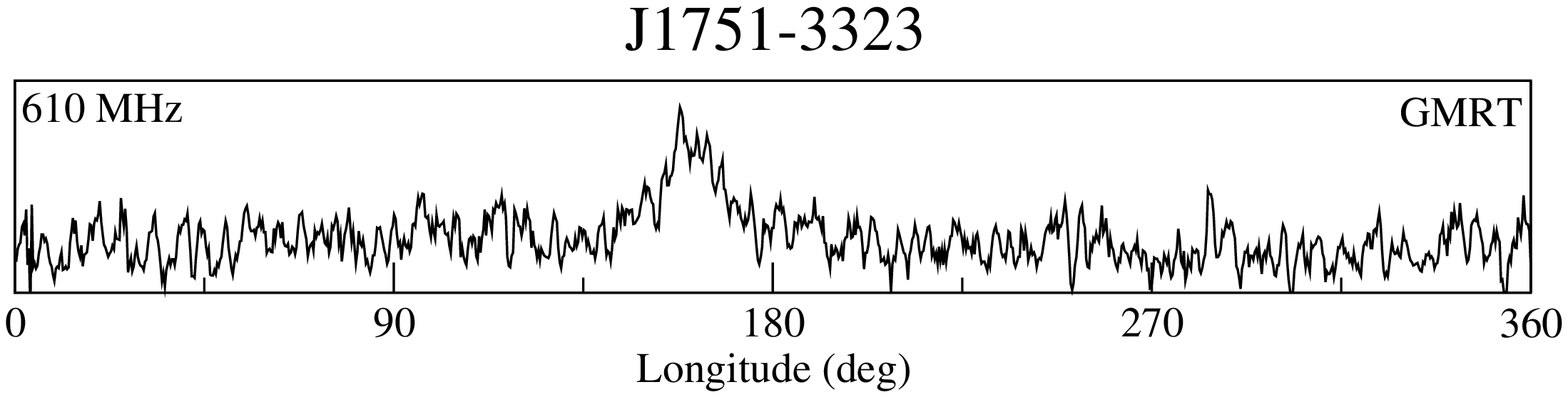}}
\resizebox{\hsize}{!}{\includegraphics[angle=-90]{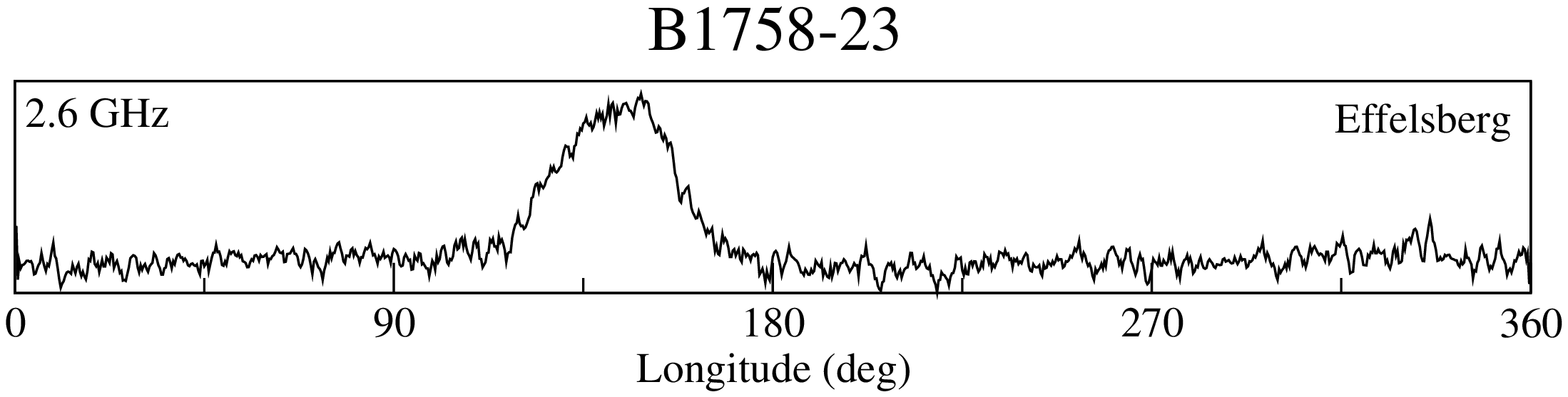}}
\resizebox{\hsize}{!}{\includegraphics[angle=-90]{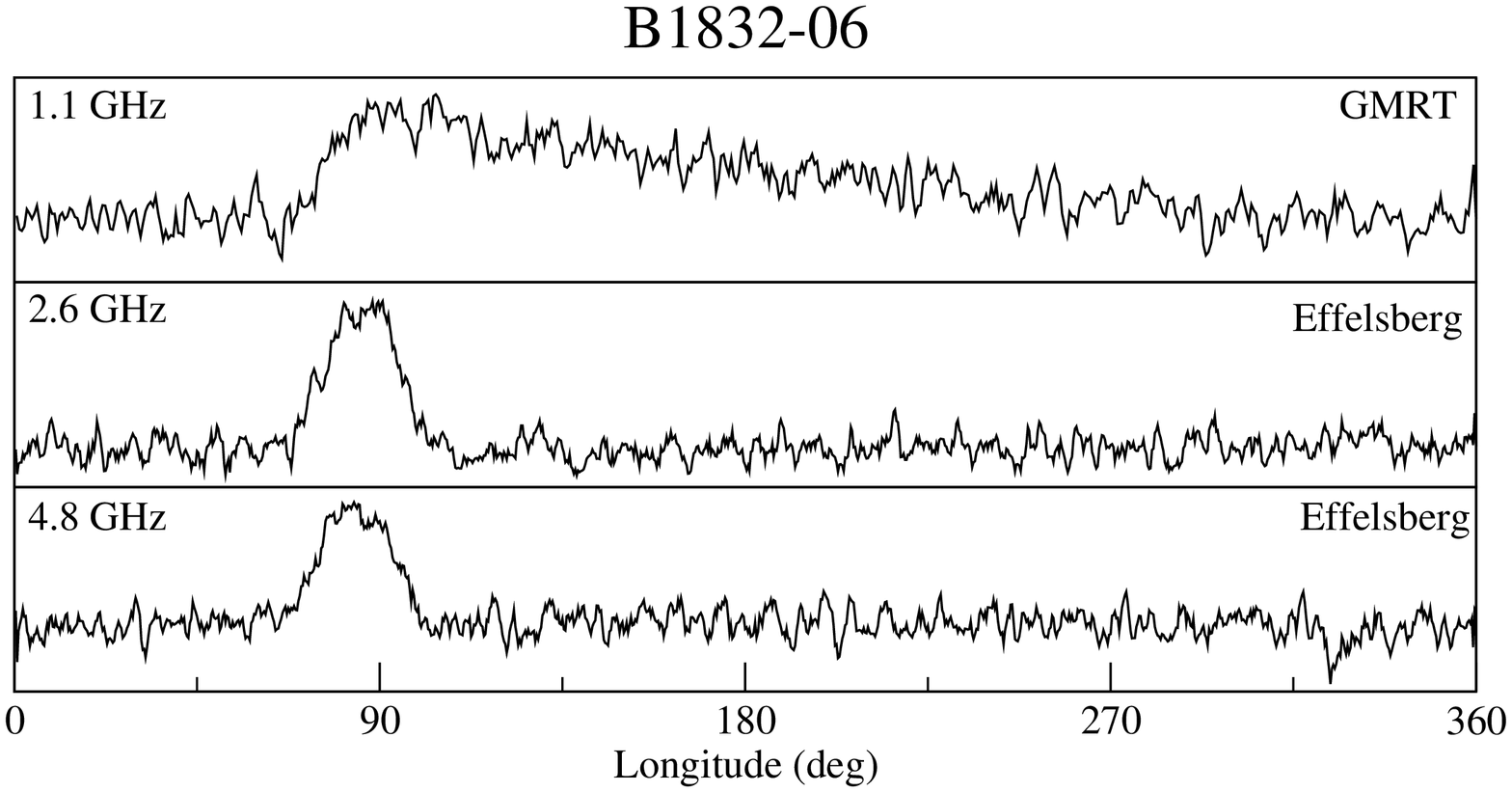}}
\resizebox{\hsize}{!}{\includegraphics[angle=-90]{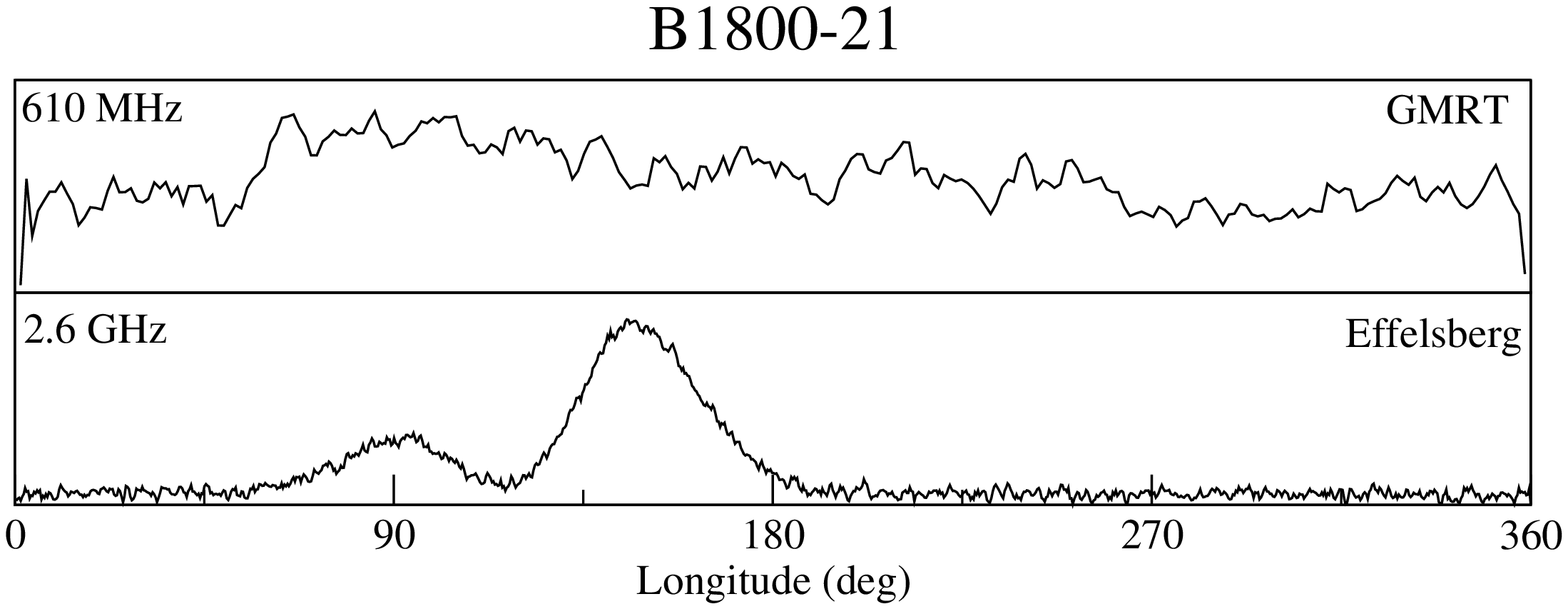}}
\caption{Appendix - Pulsar profiles.}
\end{figure}

\begin{figure}
\resizebox{\hsize}{!}{\includegraphics[angle=-90]{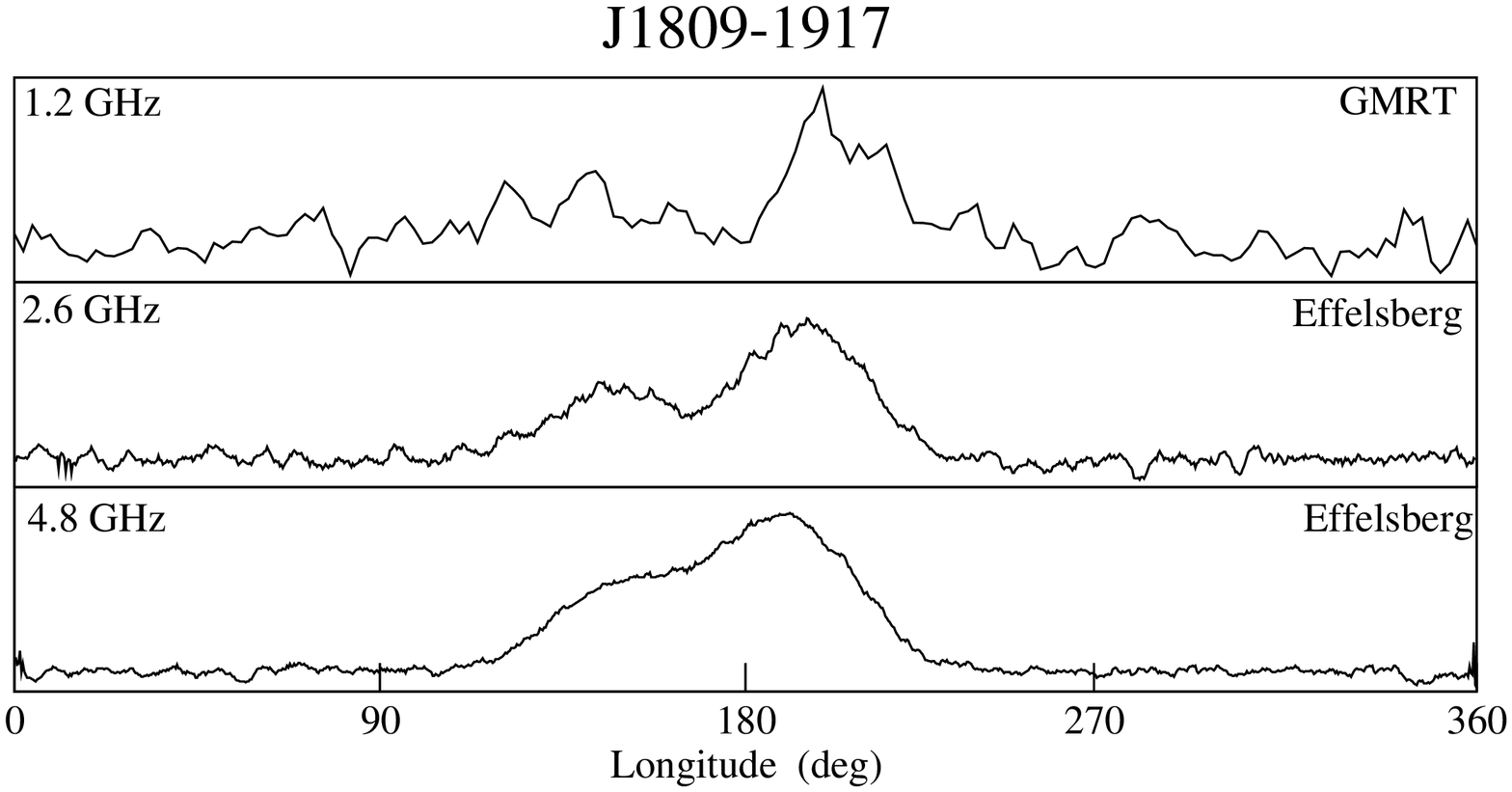}}
\resizebox{\hsize}{!}{\includegraphics[angle=-90]{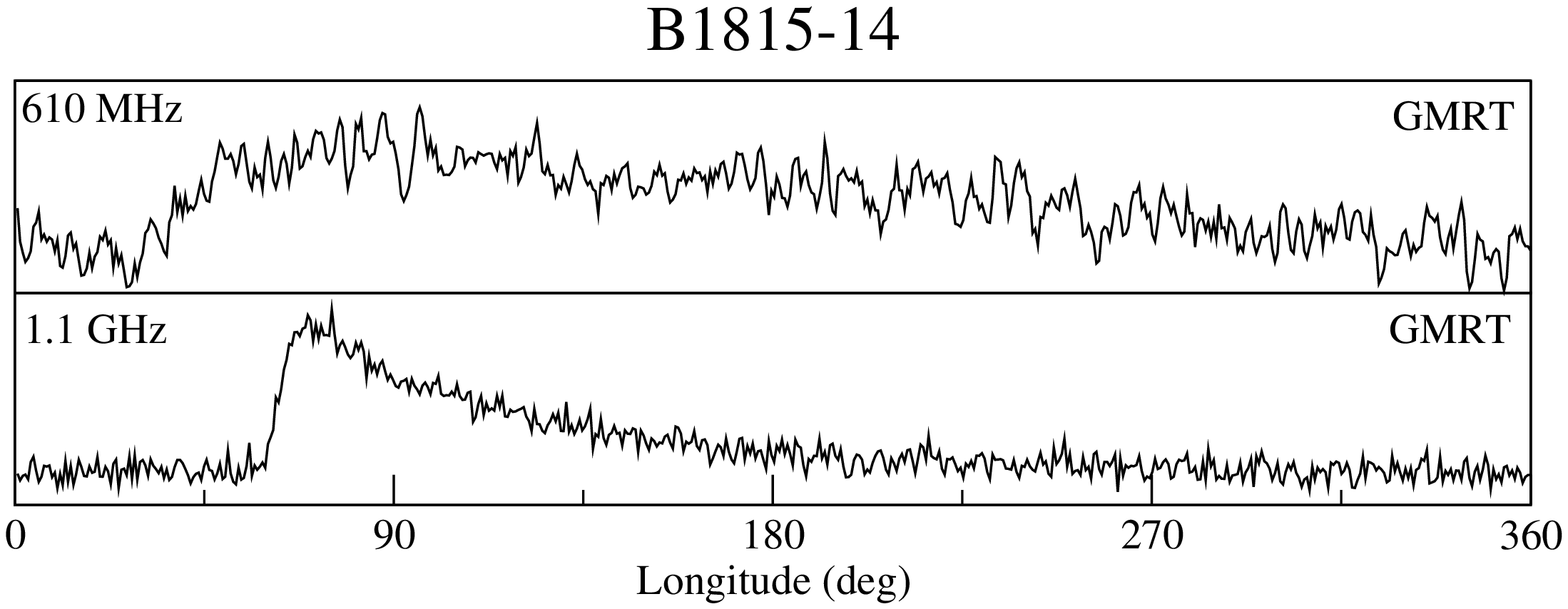}}
\resizebox{\hsize}{!}{\includegraphics[angle=-90]{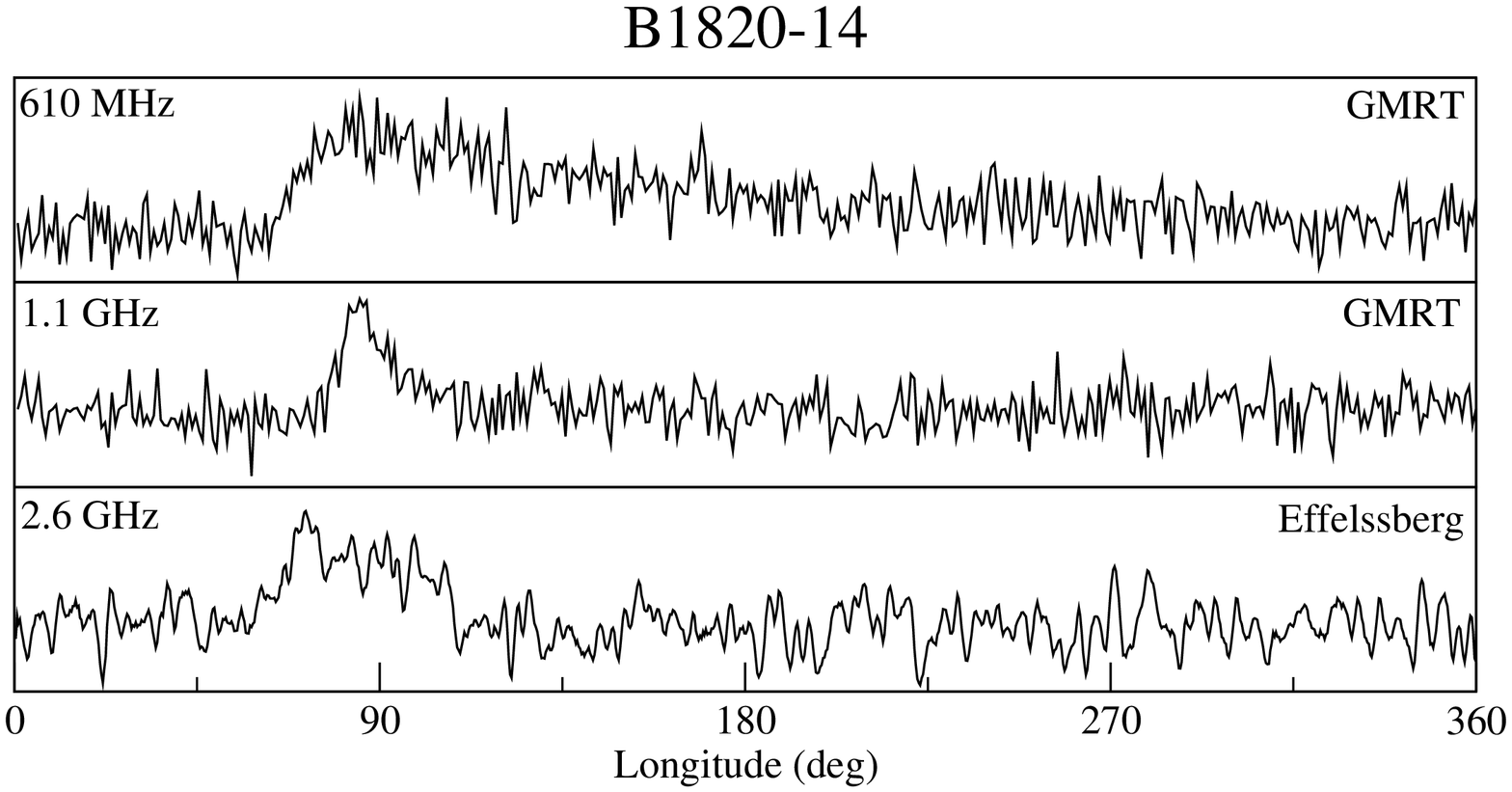}}
\resizebox{\hsize}{!}{\includegraphics[angle=-90]{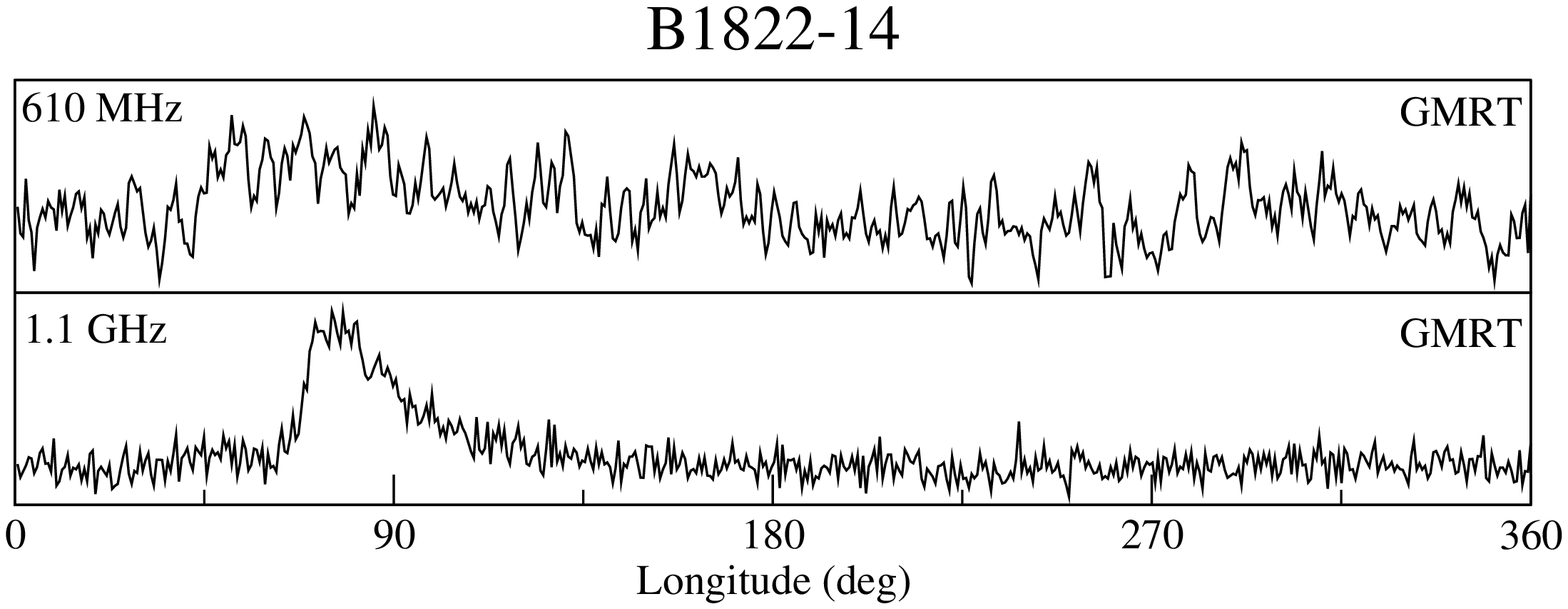}}
\resizebox{\hsize}{!}{\includegraphics[angle=-90]{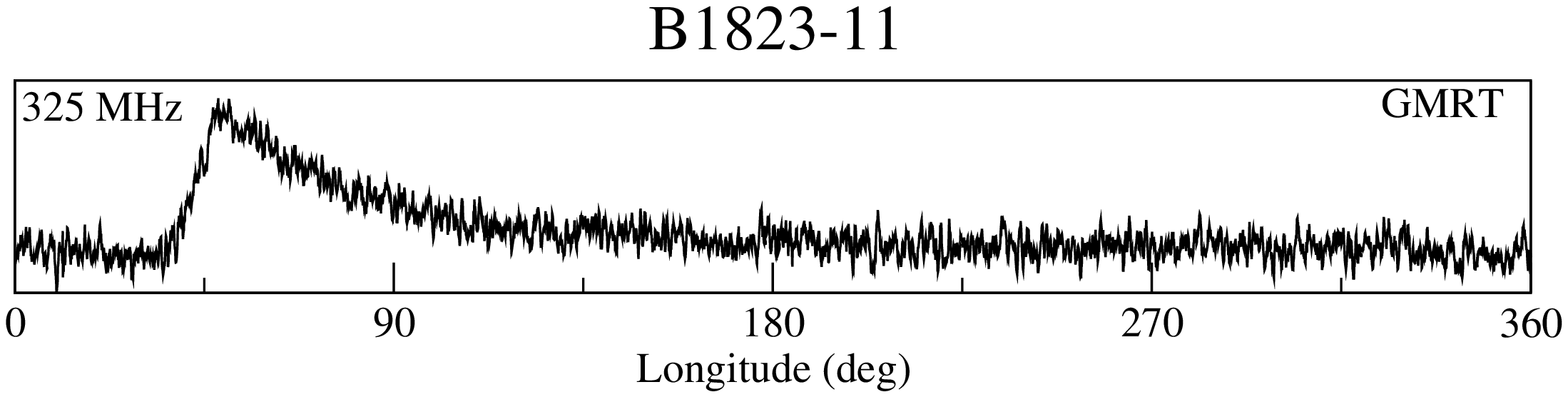}}
\resizebox{\hsize}{!}{\includegraphics[angle=-90]{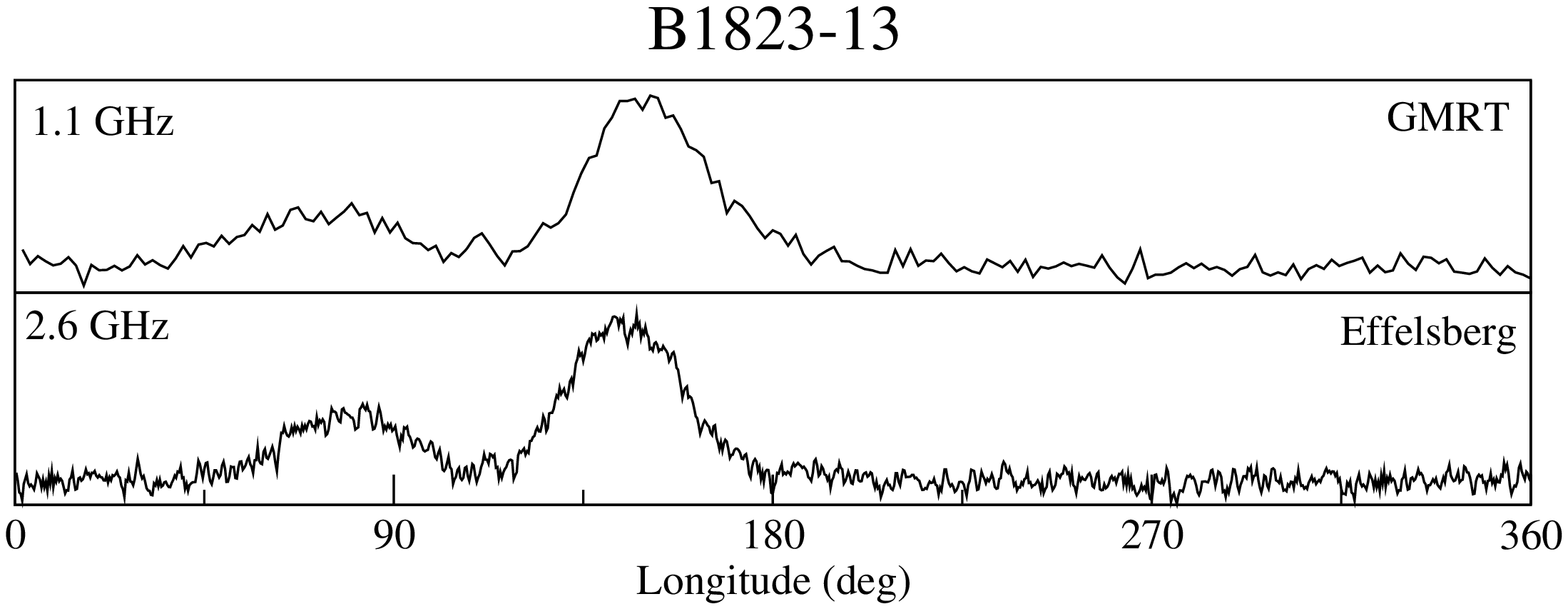}}
\caption{Pulsar profiles - continued}
\end{figure}

\begin{figure}
\resizebox{\hsize}{!}{\includegraphics[angle=-90]{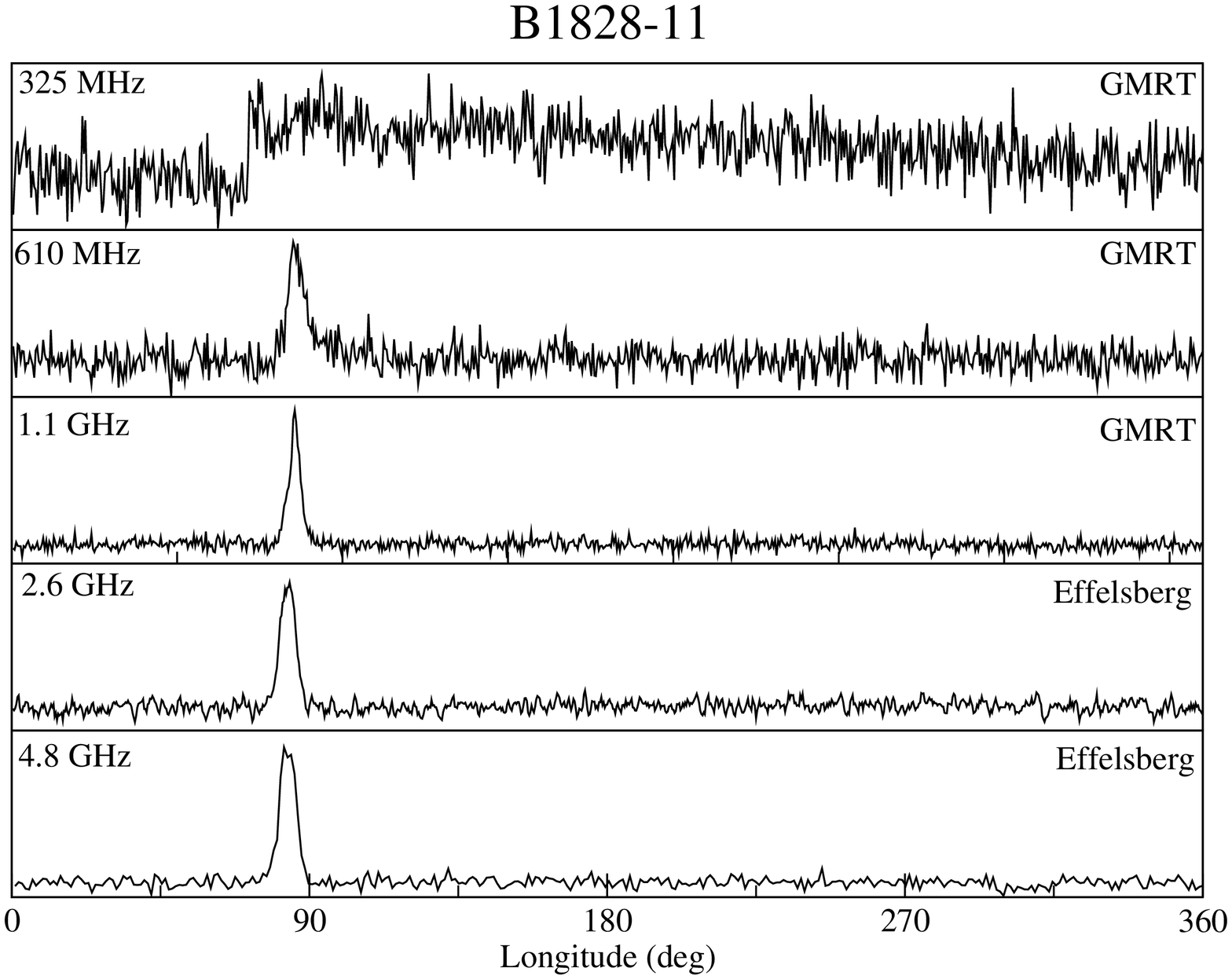}}
\resizebox{\hsize}{!}{\includegraphics[angle=-90]{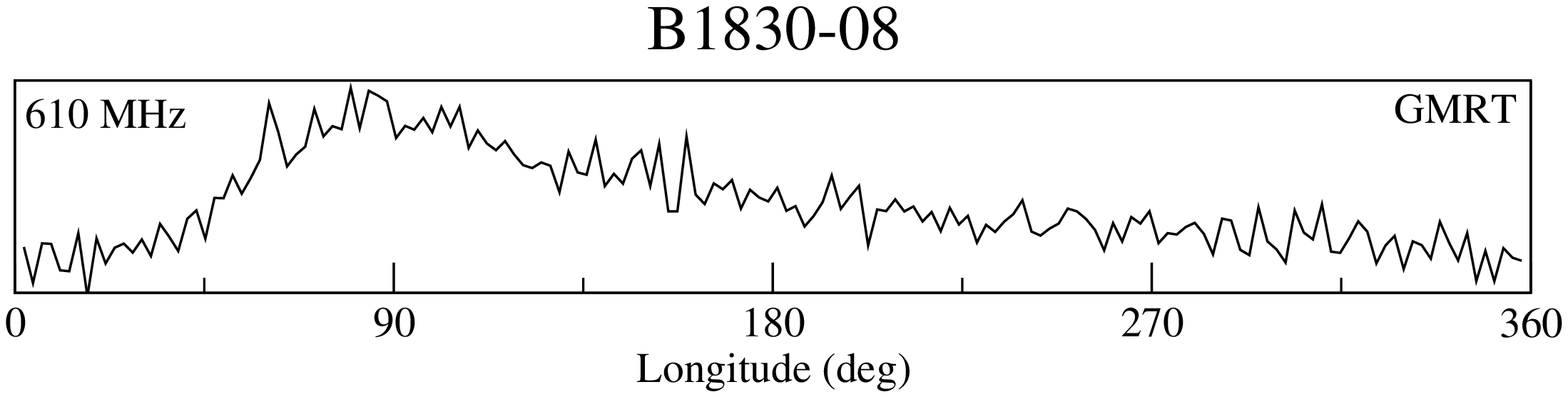}}
\resizebox{\hsize}{!}{\includegraphics[angle=-90]{b1832-06.eps}}
\resizebox{\hsize}{!}{\includegraphics[angle=-90]{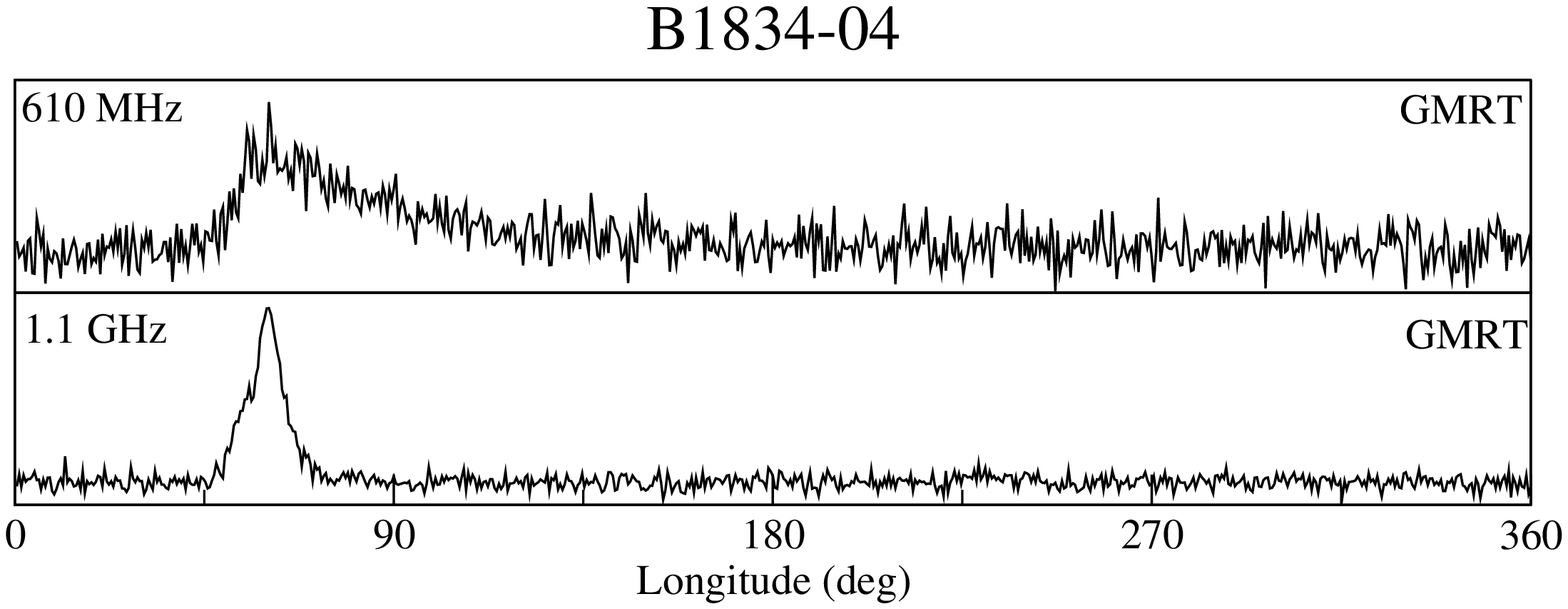}}
\resizebox{\hsize}{!}{\includegraphics[angle=-90]{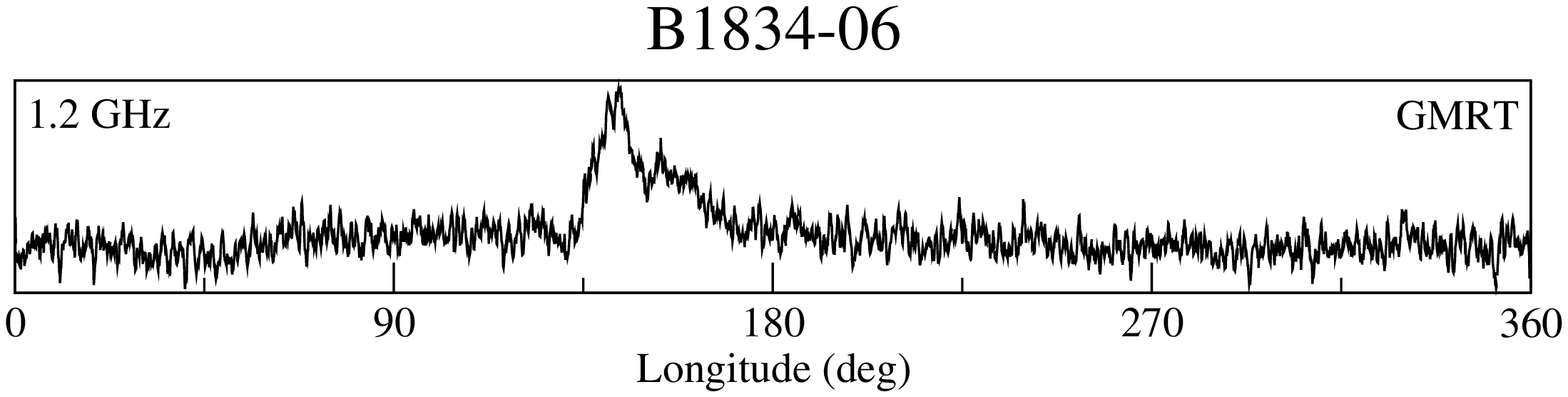}}
\resizebox{\hsize}{!}{\includegraphics[angle=-90]{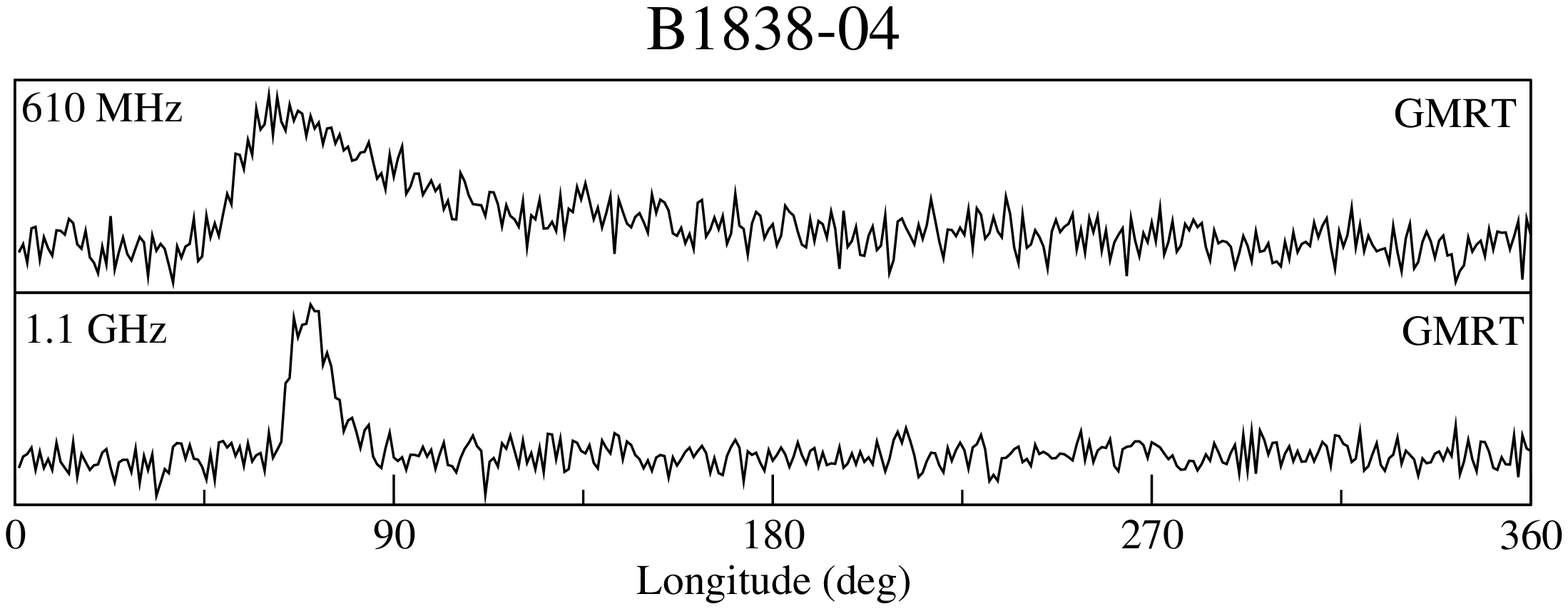}}
\caption{Pulsar profiles - continued}
\end{figure}

\begin{figure}
\resizebox{\hsize}{!}{\includegraphics[angle=-90]{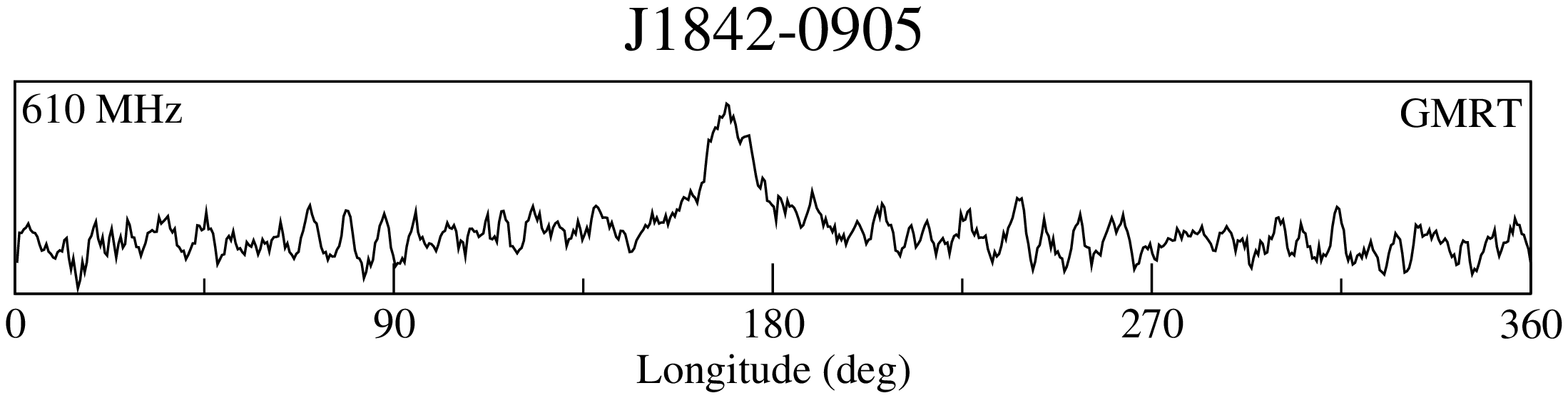}}
\resizebox{\hsize}{!}{\includegraphics[angle=-90]{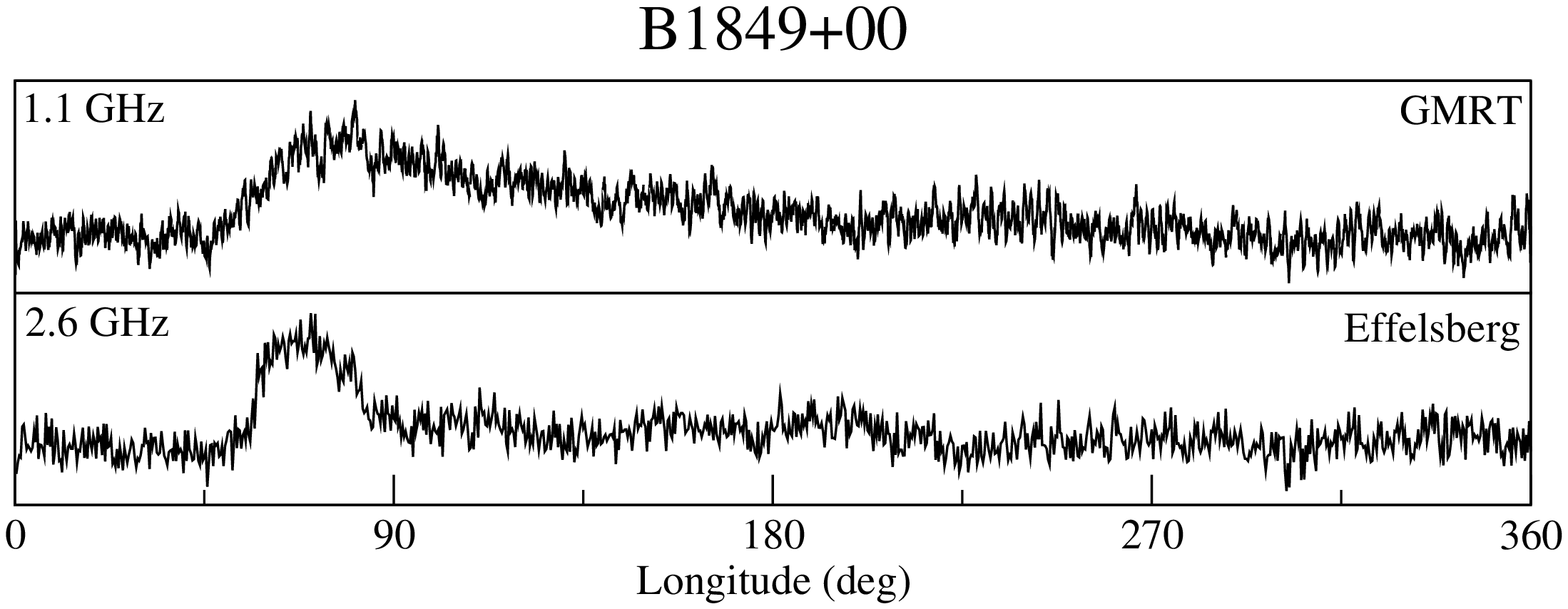}}
\resizebox{\hsize}{!}{\includegraphics[angle=-90]{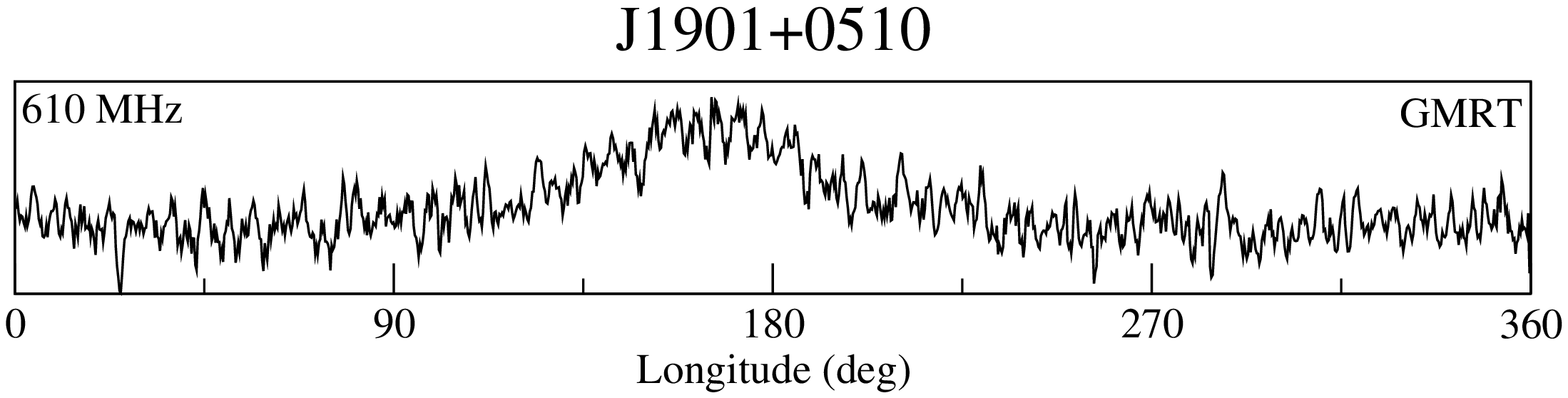}}
\resizebox{\hsize}{!}{\includegraphics[angle=-90]{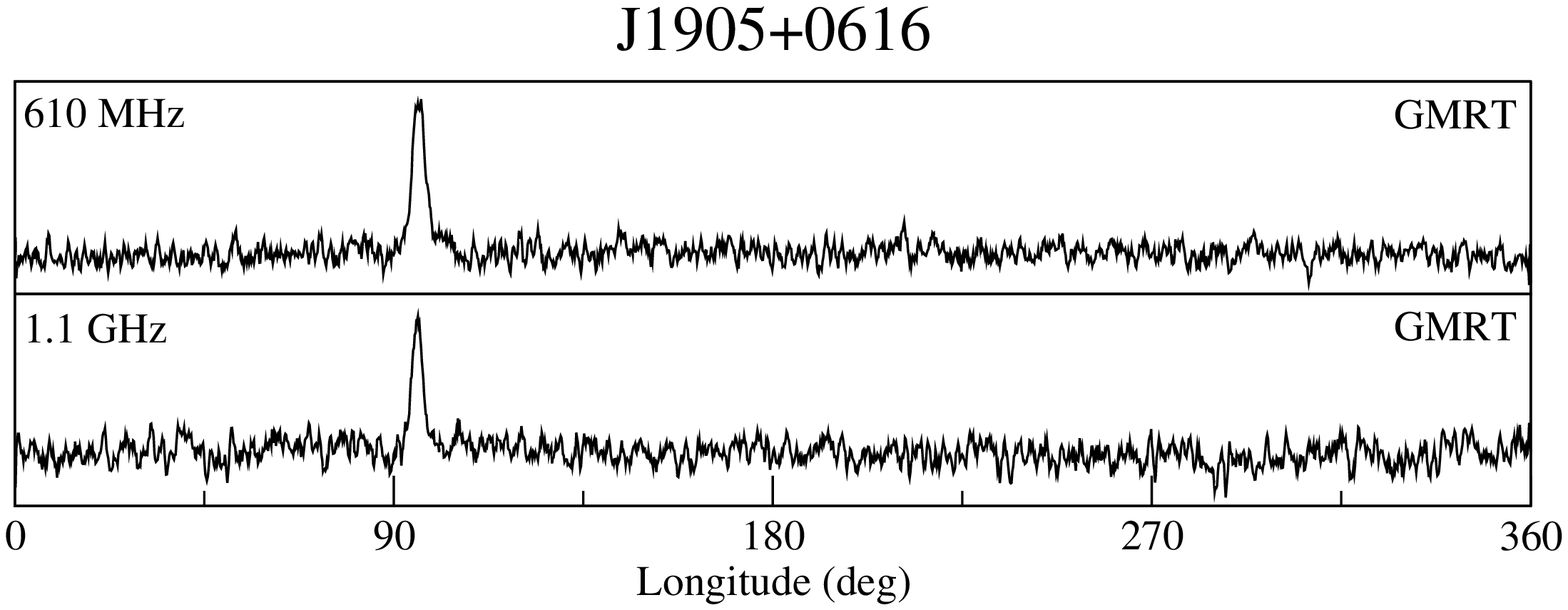}}
\resizebox{\hsize}{!}{\includegraphics[angle=0]{j1910+0728.eps}}
\caption{Pulsar profiles - continued}
\end{figure}

\end{document}